\documentclass[11pt]{article}
\usepackage[T1]{fontenc}
\usepackage[utf8]{inputenc}
\usepackage{newtxtext,newtxmath}
\usepackage[letterpaper,margin=1in]{geometry}
\usepackage{microtype}
\usepackage{graphicx}
\usepackage{float}
\usepackage{booktabs,longtable,array,calc}
\usepackage{etoolbox}
\usepackage[numbers,sort&compress]{natbib}
\bibpunct{(}{)}{,}{n}{}{,}
\usepackage{caption}
\usepackage{xurl}
\usepackage[hidelinks]{hyperref}
\usepackage{doi}
\usepackage{setspace} 

\newcommand{\real}[1]{#1}

\AtBeginEnvironment{longtable}{\footnotesize\setlength{\tabcolsep}{3pt}}
\title{\textbf{Whose Judgments Count? Representation Gaps in Crowdsourced Content Moderation Produce Unequal Protection from Perceived Toxicity}\thanks{We thank comments from Peter Bearman, Kinga Makovi, Philipp Brandt, Ryan Hagen, Alix Rule, Daniel Tadmon, Sam Donahue, Barum Park, Simone Zhang, Andy Zhao, and the members of the Networks and Context lab. We sincerely appreciate Deepak Kumar for sharing the raw annotation data with us. An earlier version of the paper was presented at the 2025 American Sociological Association Annual Meeting. This research was supported by the National Science Foundation (\# 2335815). All data and code to replicate the results will be made available upon publication of this paper. Please direct your correspondence to Zhaodi Chen \href{mailto:zc20@iu.edu}{zc20@iu.edu} or Byungkyu Lee \href{mailto:bklee@nyu.edu}{bklee@nyu.edu}}}
\author{Zhaodi Chen\\Department of Sociology\\University of South Carolina \and Byungkyu Lee\\Department of Sociology\\New York University}

\date{}

\begin{document}

\thispagestyle{empty}

\maketitle

\begin{abstract}
Content moderation is a central form of digital governance, yet people disagree over what content should be removed from shared online spaces. While platforms aggregate human judgments to build moderation systems, it remains unclear how this process shapes which users are protected from content they perceive as toxic. We address this gap by combining large-scale judgment data with counterfactual simulations that trace how the demographic composition of moderator pools shapes the distribution of protection across users. Applying this framework to removal judgments from 16,221 U.S. respondents evaluating 102,463 comments from Twitter, Reddit, and 4chan, we find demographic heterogeneities in moderation demand. We further reveal a consistent pattern of in-group protection: reductions in perceived toxicity accrue disproportionately to users who share the demographic identities of the moderator pool. Crucially, moderator pools that mirror the demographic composition of self-identified moderators on Prolific widen these disparities relative to a nationally representative baseline, while even fully representative pools fail to ensure equal protection: Black and LGB users remain underprotected unless they are represented well beyond their population share. These findings show that unequal protection from perceived toxicity can arise structurally from the aggregation of stratified removal standards, making the demographic composition of moderation inputs a key determinant of who is protected online.
\end{abstract}


\clearpage
\doublespacing 
\noindent
Open and inclusive public discourse is central to a functioning democracy, and for billions of people, such discourse now takes place on social media \citep{GottfriedPark2025,Habermas1991}. The platforms hosting these conversations wield immense power through content moderation, a form of de facto governance that defines the boundaries of acceptable speech and shapes users' online experiences, political dialogue, and access to information \citep{Gillespie2018,KozyrevaEtAl2023,Roberts2019,ChenHan2024,LeeEtAl2024,MekacherEtAl2023,OhDowney2025}. However, beyond categories of clearly illegal content, much of what moderation governs is contested: people disagree on what counts as toxic, harmful, or unacceptable content \citep{KozyrevaEtAl2023,ParkerRuths2023,PradelEtAl2024}. This raises a question of representative legitimacy: when members of a community hold different standards of acceptable speech, whose standards does the system enforce? And how do such choices distribute protection across communities? \citep{Huang2025,PanEtAl2022,MartelEtAl2025} If the adopted standards reflect a narrow demographic slice of society, some communities may be left under-protected from harmful speech.

Two recent developments have intensified this challenge by elevating aggregated human judgment to a core mechanism of online speech governance. First, platforms are increasingly using AI and automated tools to moderate \citep{GorwaEtAl2020}. Although often treated as technical solutions, these systems rely on human judgments: crowd workers label content used to train classifiers, professional moderators make removal decisions, ordinary users flag content for review \citep{Roberts2019,ChatterjeeEtAl2025,KapaniaEtAl2023,SapEtAl2022}. These socially generated inputs become the raw material for algorithmic enforcement, which can reproduce biases present in the underlying data \citep{SapEtAl2022,Davidson2026,SantyEtAl2023,SapEtAl2019}. Second, platforms are reconfiguring \emph{who} supplies such judgments. Under social and political pressure, platforms are increasingly shifting from centralized professional oversight to decentralized, community-driven models \citep{AugensteinEtAl2025,Seering2020,SlaughterEtAl2025,ZhaoHobbs2025}. For instance, in 2025, Meta replaced its professional fact-checking partnerships with community moderation amid concerns about reduced protections for vulnerable groups \citep{Kaplan2025,OversightBoard2025,Wihbey2025}, and X's 2022 moderation rollback was followed by a measurable rise in hate speech \citep{HickeyEtAl2025}.

As platforms scale governance by aggregating human judgments, moderation becomes vulnerable to a representational problem: when the judgments entering the system disproportionately reflect dominant-group standards, harms experienced by other communities may go under-recognized \citep{KapaniaEtAl2023,GordonEtAl2022,LiuEtAl2022}. Existing research shows this risk in concrete domains. For example, posts describing personal experiences of racial discrimination are disproportionately flagged as toxic by algorithms and users, and witnessing such suppression reduces Black users' sense of belonging \citep{LeeEtAl2024}. Demographic representation offers one potential answer to this problem \citep{SantyEtAl2023,DavidsonEtAl2019}. Research has shown that representative processes can enhance users' perception of moderation legitimacy \citep{PanEtAl2022,MartelEtAl2025}. Related work on machine learning systems has similarly challenged the assumption that subjective tasks like toxicity detection have a single ground truth: annotator identity shapes the datasets and models built from them \citep{SapEtAl2022,SantyEtAl2023,GoyalEtAl2022,KumarEtAl2021,XuJurgens2026}, and the composition of annotator pools can materially change classification outcomes \citep{GordonEtAl2022,LiuEtAl2022}. These insights, however, have focused primarily on annotation, model design, or perceived legitimacy, rather than examining how the demographic composition of moderation inputs affects outcomes for different communities. Answering this question is essential for understanding why representation in online governance matters, when it improves equality, and where its limits lie.

In this work, we examine whether there are systematic differences in moderation demand across gender, racial, and LGB identity groups, and how the demographic composition of moderation inputs affects the overall toxicity of online discourse as perceived by these different communities. Researchers have documented public demand for restricting harmful content and ideological divides in censorship attitudes \citep{KozyrevaEtAl2023,PradelEtAl2024,SolomonEtAl2024}. A handful of studies have identified differences along single axes, including gender and partisanship, but these findings remain fragmented \citep{AppelEtAl2023,MunzertEtAl2025,PradelTheocharis2024}. We therefore lack a systematic, multi-identity account of how moderation demand is organized across foundational social categories such as gender, race, and sexual orientation.

Also, studying the downstream consequences of stratified moderation demand faces empirical and methodological challenges. Moderation systems generally operate through opaque pipelines: platforms rarely disclose the demographic makeup of moderation teams, the relative influence of professional moderators, crowd workers, and users, or how human judgments are weighted within automated systems, limiting direct empirical leverage \citep{Roberts2019,KapaniaEtAl2023,SuzorEtAl2019,UrmanMakhortykh2023}. In addition, online toxicity is inherently perspectival \citep{SolomonEtAl2024,AppelEtAl2023}. The same content may be benign or toxic depending on social position, lived experience, and normative standards, complicating efforts to infer equitable outcomes from observed moderation decisions alone \citep{KumarEtAl2021,EadyRasmussen2025,KenskiEtAl2020}.

Our study addresses these challenges using a two-part approach combining large-scale human evaluation and a novel simulation framework. First, we examine how demographic identity structures moderation demand using a dataset where 16,221 U.S. respondents recruited through Amazon Mechanical Turk evaluated 102,463 comments drawn from Twitter, Reddit, and 4chan \citep{KumarEtAl2021}. Each respondent assessed multiple comments and provided their removal judgement, and each comment was evaluated by multiple respondents, allowing us to compare how different social groups judge identical content. We do not treat these respondents as professional moderators applying platform-specific rules; rather, their judgments approximate the socially situated standards that enter moderation systems through labeling, reporting, and community governance. We then use respondents' removal judgments to simulate which comments would be removed under moderator pools with different demographic compositions and estimate how those removals change the toxicity perceived by each group. Rather than replicating any platform's architecture, the simulation provides a clear, interpretable test of a general principle: stratified removal standards in human-generated inputs shape moderation outcomes, whatever algorithmic or institutional machinery is layered on top.

\section*{Results}

\subsection*{Who Demands More Content Removal?}

We examine how different groups evaluate and moderate identical online content using comment-fixed effects regression models. We focus on gender, race, and LGB status because these identities are the central axes of online identity-based hostility and harassment \citep{HawkinsEtAl2023,Sobieraj2018,Vogels2021} and common benchmarks for assessing algorithmic fairness \citep{BarocasSelbst2016,StarkeEtAl2022}. Our models additionally controlled for age, education, and political affiliation (full model results see SI, Table S2).

As shown in Figure 1, we find systematic group disparities in moderation demands (the direct request to remove content). Specifically, women have a greater preference for content moderation compared to men (predicted probability 0.33 vs. 0.30). The most pronounced differences emerge across racial categories. We find consistent high-demand patterns across racial minorities compared to White participants. Asian participants exhibit the highest demand for moderation (0.39), followed by Black (0.35) and Hispanic participants (0.33). Similarly, LGB respondents have higher probability of demanding content removal compared to non-LGB participants (0.33 vs 0.31). These results demonstrate a baseline of stratified moderation demand: groups susceptible to identity-based marginalization, such as women, racial minorities, and LGB individuals, are generally more likely to seek intervention when evaluating the same content. These differences in moderation demand are not reducible to differences in perceived toxicity: groups also diverge in how toxic they rate identical comments, and some high-demand groups (e.g., women, Asian respondents) perceive no more toxicity than their counterparts (SI, Table S2 and Figure S1).

\begin{figure}[htbp]
\centering
\includegraphics[width=0.96\textwidth,height=0.82\textheight,keepaspectratio]{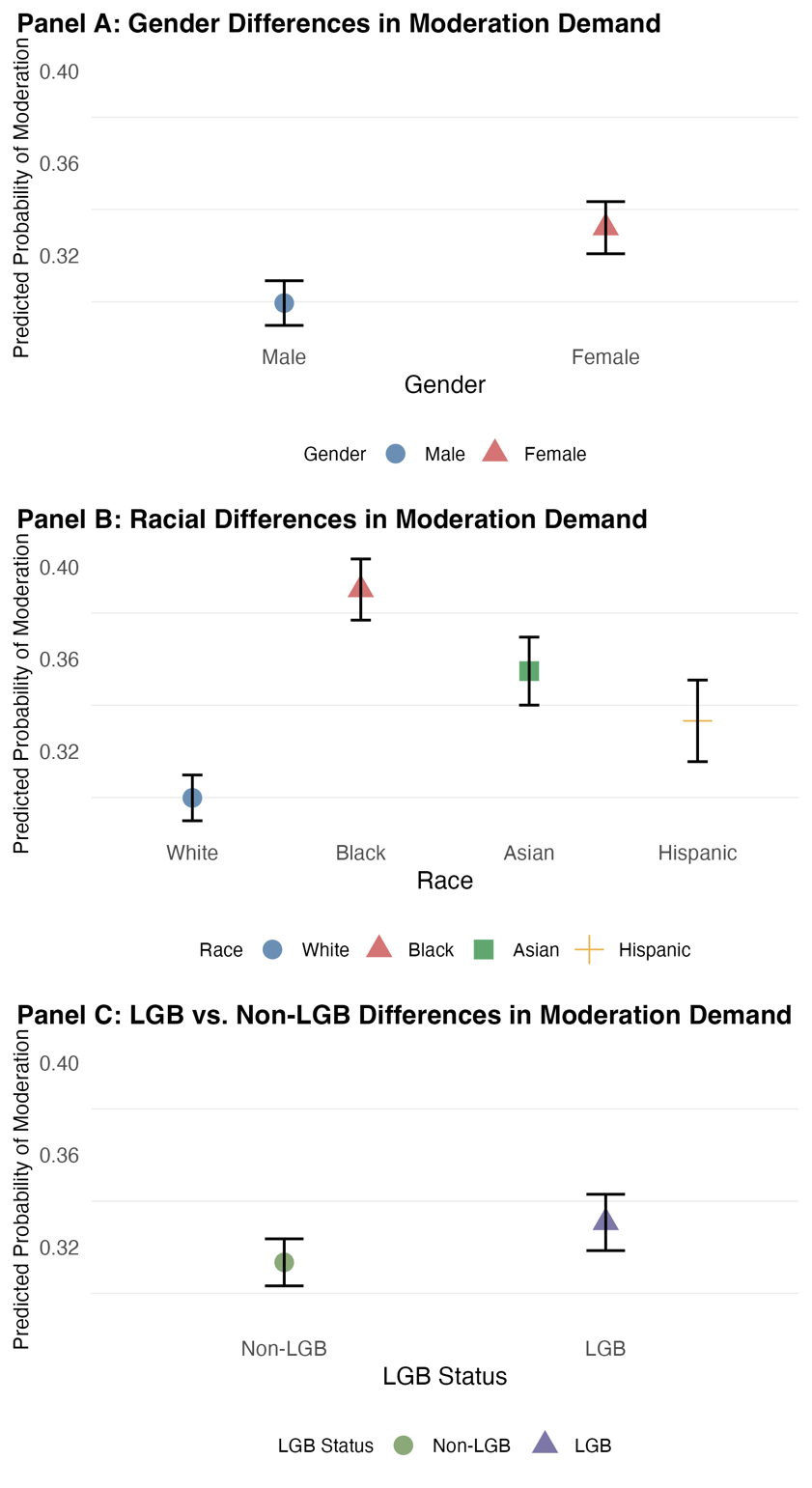}
\caption{\textbf{Group Differences in Predicted Probability of Moderation Demand.} Predicted probabilities by gender, race and LGB status, based on Model 1 (SI, Table S2). Error bars represent 95\% confidence intervals.}
\label{fig:1}
\end{figure}

\subsection*{Does Moderator Composition Change Who is Protected?}

We next trace how socially patterned moderation demand shapes moderation outcomes for different groups. A comment that most women would remove but most men would allow likely survives a mostly-male moderation team, and inequalities of this kind should be widest where removal standards divide most sharply and the moderator pool is most skewed. We examine this by simulating moderation teams of varying demographic composition. In each run, we first draw a 20-person moderator-team from our respondents under the target composition, then consider the comments rated by at least one team member. For each such comment, one member who rated it is randomly selected, and that person's recorded judgment determines removal. We then score the comments left visible after moderation against each group's own toxicity ratings. The outcome is a group's toxicity reduction: the share of perceived toxicity that moderation eliminates.

Figure 2 presents how group-specific toxicity reduction changes as the moderator pool's demographic composition varies. Across every axis, we identify a consistent in-group protection effect. That is, increasing a group's representation in the moderator pool disproportionately reduces the toxicity that group perceives in the surviving discourse. Throughout, ``protection'' refers to this reduction in group-perceived toxicity; the term describes the alignment between moderator and user identities, not moderators' motives or the targets of the comments. These gains are often accompanied by smaller reductions for others.

\begin{figure}[htbp]
\centering
\includegraphics[width=0.96\textwidth,height=0.82\textheight,keepaspectratio]{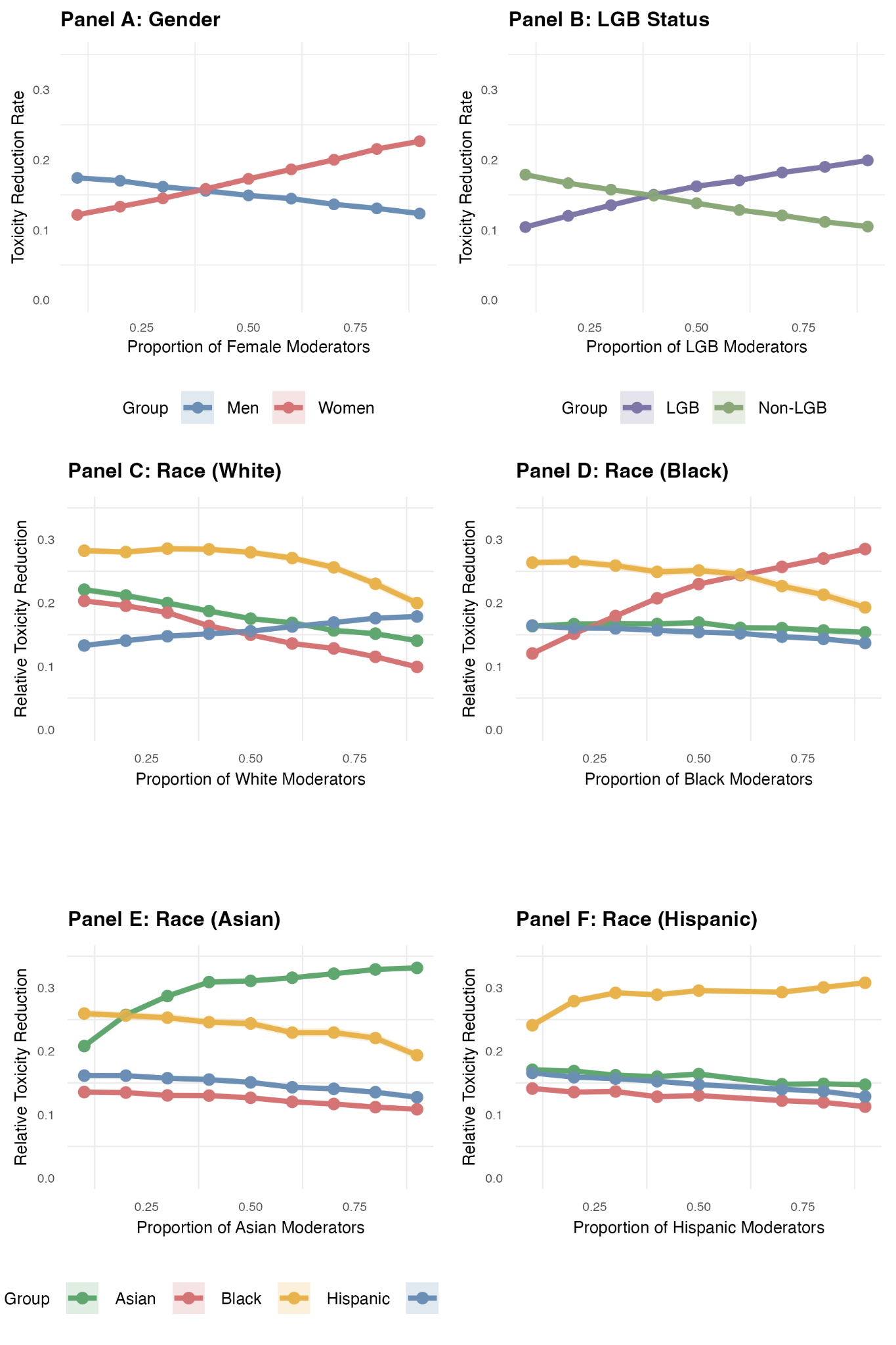}
\caption{\textbf{Simulated Effects of Varying Moderator-Pool Demographic Composition on Relative Toxicity Reduction Across User Groups.} Each panel shows group-specific toxicity reduction as the focal group's representation of the moderator pool increases. Panel A: female representation, outcomes for women and men. Panel B: LGB representation, outcomes for LGB and non-LGB users. Panels C--F: White, Black, Asian, and Hispanic representation, respectively, outcomes for four racial groups. Error bands are 95\% confidence intervals across the 2000 simulation runs per target proportion.}
\label{fig:2}
\end{figure}

In Panel A, as male representation rises from 10\% to 90\%, women's toxicity reduction declines from 22.5\% to 12\%, a nearly 50\% drop. Meanwhile, men's protection peaks under male-dominated teams. Toxicity reduction for men thus often entails preserving content that women perceive as toxic. Panel B shows the same pattern by sexual identity. In pools dominated by non-LGB moderators, LGB users experience lower toxicity reduction than non-LGB users (10\% vs. 18\%, Panel B). As LGB representation increases, protection for LGB users rises and surpasses non-LGB outcomes after LGB representation exceeds approximately 40\%. This threshold is far above the group's population share. Thus, even proportional representation may not be enough to equalize protection for numerically smaller groups whose removal standards differ from the majority.

Panels C--F reveal similar patterns. For example, increasing White representation reduces more toxicity for White respondents but reduces other groups' protection. Increasing Hispanic representation raises Hispanic protection from 25\% to 30\% with comparatively little change for other groups. Crucially, a concerning pattern emerges regarding the systemic vulnerability of Black users across varying moderator compositions. Black users reach nearly 30\% toxicity reduction under Black-dominated pools. But they receive the least protection in almost every other composition. That is, the forms of toxicity Black users perceive as most toxic appear structurally under-removed by out-group moderators. This suggests that without deliberate attention to cross-group protection, demographic shifts in moderator composition can inadvertently create protection gaps for marginalized communities.

The pattern is not limited to the single-decision rule. To test whether it persists when several moderators' judgments are combined for the same comment, we repeated the simulation under a set of alternative decision rules: majority of three, majority of five, supermajority of five, removal if any of five moderators flags the comment, majority of ten, and majority of twenty (SI, Table S3). Because respondents did not rate every comment, implementing these rules requires estimating unobserved respondent-comment removal decisions. We used a calibrated imputation procedure that preserves observed decisions and estimates removal probabilities for unrated comments (Materials and Methods). The results are consistent across decision rules. Under the majority-of-five rule, for example, increasing a group's representation shifts the toxicity-reduction gap toward that group: the focal group's relative toxicity reduction rises while the comparison group's reduction generally declines or grows more slowly. LGB respondents remain less protected than non-LGB respondents until LGB representation reaches roughly 70\%, and Black respondents remain less protected than White respondents until Black representation reaches roughly 60\% (SI, Figure S2). Full rule-specific results are reported in the SI, Table S4.

\subsection*{Who is Left Under-Protected Under Observed Moderator Demographics?}

Having isolated each demographic axis, we next ask how these dynamics play out under empirically grounded baselines. We construct moderator pools matching four benchmark populations: (1) the general U.S. population based on the General Social Survey (GSS) (the general-population pool), (2) demographically representative Prolific respondents who self-reported moderation experiences (the moderation-experience pool), (3) Prolific respondents without such experience (the no-experience pool), and (4) the published demographics of Reddit moderators, a prominent community-moderation setting (the Reddit-matched pool) \citep{WangEtAl2022,Matias2019,LiEtAl2022}. Compared to the U.S. population, the observed benchmark pools are younger, more educated, more male-dominated, overrepresent White and Black users, and have higher LGB representation (SI, Table S5). The Reddit-matched pool is especially male dominated. Because published Reddit moderator data cover gender, age, and education but not race or sexual identity, the Reddit-matched pool is specified on those margins only (Materials and Methods). We do not read these simulations as estimates of any platform's actual outcomes. By holding platform rules and enforcement processes constant, these simulations isolate how the composition of the judgment pool, by itself, shifts the distribution of protection across groups.

Figure 3 presents the results. Panel A demonstrates a clear shift in gender-based moderation outcomes contingent on moderator demographics. Under the Reddit-matched (predominantly male) pool, men experience greater toxicity reductions than women (15.0\% vs. 12.1\%). Under the gender-balanced general-population pool, the advantage reverses toward women (17.5\% vs. 14.9\%). The moderation-experience and no-experience pools fall between these endpoints and show smaller female-favorable gaps.

\begin{figure}[htbp]
\centering
\includegraphics[width=0.96\textwidth,height=0.82\textheight,keepaspectratio]{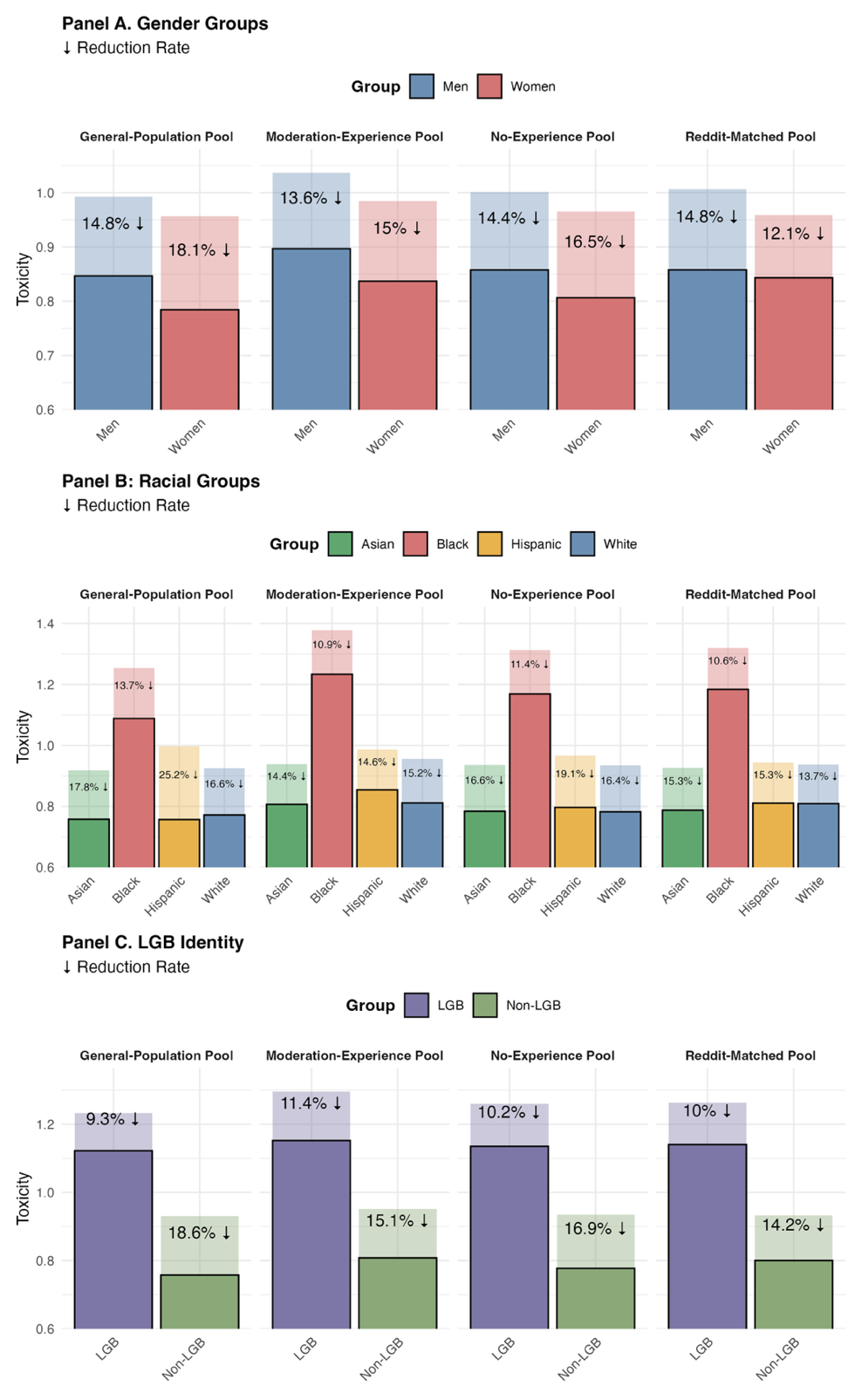}
\caption{\textbf{Simulated Relative Toxicity Reduction Across Four Empirically Grounded Scenarios (General Population, Moderation-Experience, No-Experience, and Reddit-Matched Pools).} Panel A: pre- and post-moderation toxicity levels for women and men. Panel B: for four racial groups. Panel C: for LGB and non-LGB user groups. Shaded background bars show mean pre-moderation toxicity. Solid bars show post-moderation toxicity. Annotations denote relative toxicity reduction. Values are means across 2000 runs per scenario. The Reddit-matched pool is specified on gender, age, and education margins only, as the Reddit data do not report race or sexual identity (Materials and Methods).}
\label{fig:3}
\end{figure}

Panel B shows that shifting from the general-population pool to the moderation-experience and Reddit-matched pools lowers protection for most racial groups, but the losses are unequal. Hispanic users' protection drops by nearly half (26.6\% to 14.6--15.3\%). Asian users, similarly underrepresented, see a smaller erosion (16.7\% to 14.4--15.3\%). White users' toxicity reduction also declines somewhat despite their high representation. This pattern may reflect an intersectional trade-off: male overrepresentation, associated with lower removal demand, can offset the advantage of White representation. Most strikingly, Black participants consistently receive the lowest toxicity reductions across all moderator compositions (10.6\% to 13.5\%), suggesting systematic under-protection regardless of moderator demographics. The persistent under-protection indicates a fundamental structural disadvantage: neither the representative pool nor the observed benchmark compositions prioritize the specific form of toxicity perceived by Black users.

Panel C shows a parallel structural ceiling for sexual minorities. LGB users experience less toxicity reduction than non-LGB users in every benchmark (9.3--11.4\% versus 14--18\%). This persistent disparity illustrates how the in-group protection identified earlier may manifest in empirically observed pool compositions: as Figure 2 showed, effective protection for LGB users requires LGB representation well above 40\%. This threshold far exceeds most plausible real-world pools. Because LGB users remain a numerical minority even in ``overrepresented'' pools, the content they disproportionately perceive as toxic remains structurally deprioritized in the simulation.

Together, these simulations show that moderator-pool demographics can redistribute protection across groups in practically consequential ways. Compared with a nationally representative pool, empirically grounded pools consistently widen protection gaps across demographic axes. Making pools proportionally representative narrows most of these gaps but does not close them all: Black and LGB users remain least protected even under representative conditions, revealing the limits of descriptive representation. These patterns are robust to the alternative decision rules (SI, Table S6 and S7).

\section*{Discussion}

Content moderation increasingly structures public discourse, yet its core task remains fundamentally contested: determining whether and when toxic content warrants intervention \citep{KozyrevaEtAl2023,ParkerRuths2023,PradelEtAl2024}. Our research contributes to this debate by providing a multi-identity analysis linking demographic differences in moderation demand to the distribution of protection produced by differently composed judgment pools. We find that women, racial minorities, and LGB individuals are more likely than their counterparts to judge identical comments as warranting removal. Our simulations demonstrate that these differences, when aggregated through moderation processes, could be translated into unequal protection across groups.

A central implication of these findings is that the demographic composition of the moderation judgment pool can become a key source where unequal online experiences arise. Platforms routinely decide who labels training data, who moderates communities, whose reports trigger review, and whose judgments are used to train or evaluate automated systems. These choices are often treated as technical or operational features of moderation infrastructure. Our results show that they have distributional consequences: changing whose removal judgments enter the system changes whose perceived toxicity is reduced. As platforms increasingly rely on AI-driven and community-based moderation trained on aggregated human judgments, inequalities in protection can arise as a structural byproduct of aggregating stratified standards \citep{ZhaoHobbs2025}. We should therefore treat the composition and aggregation of human inputs, ranging from moderator teams and annotator pools to user-reporting populations and community-note contributors, as a central object of digital governance.

Our simulations identify one consequence of this governance choice: demographically skewed pools can channel greater protection toward the dominant group, suggesting a pattern of in-group protection. A useful lens for interpreting this pattern comes from research in social psychology on \emph{selective empathy}, the tendency to direct empathic attention unevenly across parties in a moral dispute \citep{HassonEtAl2022,KiviatKnight2026}. Content moderation is inherently relational, adjudicating between the targets of harmful speech, the speakers who may be sanctioned, and the broader community whose norms are at stake. Because empathic perspective-taking is socially patterned by identity and social distance, demographically homogeneous pools may more readily ``see'' and prioritize harms that resonate with their own group experiences while treating other harms as less salient, less credible, or more tolerable. In other words, individuals may agree that content is harmful yet diverge in whether it merits removal depending on which parties they empathize with and which tradeoffs they consider legitimate.

The findings also clarify both the value and the limits of descriptive representation---the principle that a governing body should demographically mirror the population it serves---in content moderation. We find that representation helps: more representative judgment pools reduce protection gaps. However, we also identify a structural ceiling to this approach as Black and LGB users are persistently under-protected even under representative conditions. This pattern suggests that certain forms of harm may remain insufficiently recognized even within diverse decision-making bodies. We need approaches that explicitly govern whose judgments the moderation system learns to emulate, rather than defaulting to majority vote that can override minority perspectives \citep{GordonEtAl2022,XuJurgens2026}. One promising direction is \emph{jury learning}, an approach that lets practitioners explicitly specify which demographic groups, and in what proportions, a classifier should learn to reflect, enabling counterfactual tests of how alternative compositions would change moderation outcomes \citep{GordonEtAl2022}.

Several limitations constrain our findings' generalizability. Our sample represents U.S. perspectives and may not capture moderation demand patterns in other countries. Our survey respondents are not professional moderators. Those working in content moderation may have different preferences, understanding, or decision-making patterns compared to the general population \citep{ChatterjeeEtAl2025,RuckensteinTurunen2020}. Our comment dataset may not capture the full range of toxic speech patterns across different platform contexts. Our simulation necessarily simplifies complex sociotechnical systems through which moderation operates, but real-world moderation involves multiple stages of human and algorithmic review, appeal processes, and contextual factors our model does not capture. Future research should examine how representational effects manifest within more complex, multi-stage workflows and across different platform governance models.

\section*{Materials and Methods}

\textbf{Primary Data.} Our primary data come from the \emph{Toxicity Perspectives} dataset \citep{KumarEtAl2021}, in which 17,280 U.S. respondents evaluated 107,260 online comments for toxicity and made subsequent moderation decisions (see SI, Table S1 for descriptive statistics). Participants were recruited via Amazon Mechanical Turk. The nonprobability sample is not nationally representative. Compared to the 2024 General Social Survey (GSS), it is younger, more educated, and contains slightly larger shares of White and Black respondents (SI, Table S5). Respondents first rated each comment's perceived toxicity from 0 (``Not at all toxic'') to 4 (``Extremely toxic'') and then chose whether it ``should be allowed'', ``depends on the context,'' or ``should be removed.'' Our primary outcome codes only ``should be removed'' as removal. This conservative specification treats conditional approval as insufficient for removal. Robustness analyses use the broader alternative coding.

The analytical sample contains 16,221 adult respondents, 102,463 comments, and 501,540 ratings. On average, each comment received about five ratings. We exclude respondents with missing responses on race, education, age, or LGB status, as well as nonbinary respondents (about 1\% of the sample), whose number is too small to support the composition simulations. Comments come from Twitter, Reddit, and 4chan, capturing a broad spectrum of online discourse. Because most online content is non-toxic, the dataset uses stratified sampling to oversample comments likely to be toxic, especially those with high inter-rater disagreement \citep{KumarEtAl2021}. This design concentrates information on contested judgments rather than estimating the prevalence of toxicity on any platform.

\textbf{Additional Data Sources.} Three sources define empirical benchmarks (SI, Table S5). (a) The 2024 GSS supplies the demographic profile of the general U.S. population. (b) An original demographically representative Prolific survey (N = 1,161; March 2025) identifies respondents with and without content moderation experience (instrument in SI, Table S8). Moderation experience is self-reported and spans volunteer, community, and employment contexts; the resulting sample characterizes the broad population of people who perform moderation tasks---the population most relevant to crowdsourced and community-driven moderation---rather than professional platform moderators specifically. (c) Published survey data provide the profile of Reddit moderators \citep{WangEtAl2022}, an example of decentralized, community-driven moderation.

\textbf{Statistical Models.} We examine group differences in moderation demand using comment fixed-effects logistic regressions of the binary removal decision on respondent characteristics (race, gender, LGB status, education, age, political affiliation; specification in SI) and report average predicted probabilities holding other covariates at observed values.

\textbf{Simulation Procedure.} Because moderation pipelines are opaque and systematic evidence on moderator demographics is scarce, directly estimating demographic effects from observed platform outcomes is difficult. We therefore develop a counterfactual simulation framework that varies the demographic composition of hypothetical moderator teams while holding other institutional factors constant. In each run, we sample a 20-person moderator pool from the analytical sample and collect the comments rated by its members. Every comment is assigned to one randomly selected member, whose recorded response determines whether it is removed. We repeat this 2,000 times with a new team each run (full procedure in SI, Figure S3). When targeting a specified composition, it is crucial to prevent demographic confounding, whereby changing one characteristic (e.g., female representation) inadvertently alters correlated attributes. To avoid this, we apply a raking procedure using demographic margins from a reference population (e.g., GSS for nationally representative benchmarks). The resulting weights vary the targeted demographics as intended while holding other demographic consistent with the reference population. We implement raking using the \emph{autumn} package in R.

\textbf{Alternative Decision Rules.} We repeat the simulation under six alternative decision rules that combine several moderators' judgments for the same comment: majority of 3, 5, 10, or 20; supermajority of 5; removal if any of 5 moderators flags (SI, Table S3). Because respondents did not rate every comment, these rules require estimating unobserved respondent-comment decisions. We estimate them with an annotator-conditioned model that predicts the probability that each respondent would remove each comment from the comment text, toxicity rating, a learned rater-specific embedding, and the respondent's demographic characteristics (details in SI, Supporting Text; validation in SI, Figure S4). In each run, we sample moderator pools as above, assign moderator panels to all 102,463 comments, and compute the probability that the rule removes each comment given panel members' removal probabilities. Post-moderation toxicity then weights each comment by its probability of remaining visible. The imputation expands only the moderation-decision side of the simulation. We do not impute missing group-specific toxicity ratings, so each group's toxicity outcomes use only its observed ratings (coverage diagnostics in SI, Table S9).

\textbf{Simulation Scenarios.} We simulate two types of scenarios. Counterfactual scenarios vary a single attribute (e.g., female representation from 10\% to 90\%), isolating how representation along each axis shapes outcomes. Empirically grounded scenarios match the moderator pool to the general U.S. population, Reddit moderators, and Prolific respondents reporting moderation experience versus no experience. Published data on Reddit moderators report gender, age, and education, but not race or sexual identity \citep{WangEtAl2022}. The Reddit-matched pool is therefore raked to the available margins only; its racial and LGB composition is not independently constrained but follows from the survey sample conditional on those targets.

\textbf{Outcome Measures.} Following each simulation, we assess two outcomes for each demographic group: (1) absolute toxicity reduction, the difference between mean group-perceived toxicity before moderation and among surviving comments, and (2) relative toxicity reduction, the absolute reduction divided by pre-moderation toxicity (formal definitions in SI). Reported estimates are means across 2,000 runs. Intervals are 95\% Monte Carlo confidence intervals for the mean across runs.

\textbf{Diagnostics.} To ensure simulation validity, we verify at each stage that raked weights and sampled pools match target demographic compositions, for both the main and alternative decision-rule simulations (SI, Figure S5-S7, Tables S10-S11).

\textbf{Robustness Checks.} Results are robust to two alternative specifications. (1) An alternative coding of moderation decisions that treats ``It depends on the context'' as supporting removal (SI, Figure S8-S10). (2) Models using respondents' personal filtering decisions (``I don't want to see this'' vs. ``This is fine for me to see''), capturing individual filtering rather than global removal (SI, Figures S11--S13).

\clearpage

\bibliographystyle{unsrtnat}
\bibliography{references}

\clearpage
\section*{Supporting Information}

\section*{Supporting Text}

\subsection*{Statistical Model Specification}

This section provides formal specifications of the statistical models reported in the main text. We examine group differences in moderation demand using regression models with comment-level fixed effects. By holding comment-specific factors constant (e.g., content, tone, context), these models isolate the influence of respondent characteristics on moderation decisions. We estimate logistic regression models with controls for respondent-level characteristics, including race, gender, LGB status, education, age, and political affiliations. The model is specified as follows, where $M_{ij}$ is the binary moderation decision and $\mathbf{X}_i$ is the vector of respondent characteristics:

\begin{equation}\log\!\left(\frac{\Pr(M_{ij}=1)}{1-\Pr(M_{ij}=1)}\right)=\gamma_j+\boldsymbol{\delta}\mathbf{X}_i.\end{equation}

We report average predicted probabilities, holding other covariates at their observed values.

\subsection*{Formal Definitions of Outcome Measures}

This section provides formal definitions of the two simulation outcome measures reported in the main text. Following each simulation, we assess two structural outcomes related to online discourse for each demographic group. Let $C_s$ denote the specific subset of comments assigned to the moderation team during run $s$. $T_c^{(g)}$ denote the mean perceived toxicity of comment $c$ as evaluated by group $g$, and $M_{c,s}\in\{0,1\}$ represent the binary moderation decision made by the randomly assigned moderator (where 1 indicates removal). The pre-moderation mean toxicity for group $g$ in run $s$ is defined as:

\begin{equation}\mu_{\mathrm{pre},s}^{(g)}=\frac{1}{\lvert C_s\rvert}\sum_{c\in C_s}T_c^{(g)}.\end{equation}

The post-moderation mean toxicity for group $g$ is calculated based only on the comments that were not removed during the simulation run:

\begin{equation}\mu_{\mathrm{post},s}^{(g)}=\frac{\sum_{c\in C_s}T_c^{(g)}(1-M_{c,s})}{\sum_{c\in C_s}(1-M_{c,s})}.\end{equation}

Using these baselines, we calculate two key metrics. \emph{(1) Absolute Toxicity Reduction}, defined as the difference in mean perceived toxicity before and after moderation:

\begin{equation}\mathrm{ATR}_{s}^{(g)}=\mu_{\mathrm{pre},s}^{(g)}-\mu_{\mathrm{post},s}^{(g)}.\end{equation}

And (2) \emph{Relative Toxicity Reduction}, defined as the absolute reduction divided by mean pre-moderation toxicity to capture proportional improvements in discourse:

\begin{equation}\mathrm{RTR}_{s}^{(g)}=\frac{\mathrm{ATR}_{s}^{(g)}}{\mu_{\mathrm{pre},s}^{(g)}}.\end{equation}

We compute both measures separately for each demographic group to capture how moderation-induced changes in toxicity level differ across groups. Reported simulation estimates are means across 2,000 runs. When intervals are shown, they are 95\% Monte Carlo confidence intervals for the mean simulation estimate across runs.

\subsection*{Imputation of Unobserved Moderation Decisions}

\textbf{Rationale.} The alternative decision rules combine several moderators' judgments on the same comment, but each comment was rated by approximately five respondents, so the respondent-by-comment decision matrix is largely unobserved, and observed coverage is uneven across demographic groups (Table S9). Implementing multi-moderator rules therefore requires estimates of decisions that respondents did not record. We retained observed decisions when available: wherever a respondent rated a comment, the recorded decision is used, and the model supplies only the missing cells. Group-specific toxicity ratings are not imputed; they are used only where observed (see \emph{Use in the Simulation} below).

\textbf{Model.} We estimate each respondent's probability of removing each comment with an annotator-conditioned model, following the jury-learning approach of modeling judgments conditional on who is judging \citep{GordonEtAl2022}. The model combines three inputs: the comment text, encoded by a toxicity-domain language model (HateBERT) together with the Perspective API content scores included in the dataset; a learned respondent-specific embedding; and the respondent's demographic characteristics (race, gender, LGB status, education, age, and political affiliation). We train a single model across all respondents rather than separate models per group, because the smaller groups provide too few ratings to support independent models and per-group training discards the signal shared across respondents. Predicted probabilities are calibrated so that they match observed removal patterns.

\textbf{Individual- and Group-Level Accuracy.} The model is trained on the 501,540 observed judgments and evaluated on held-out data, including comments never seen during training. It discriminates individual removal decisions with AUC-ROC = 0.84, compared with 0.74 for a text-only model and 0.66 for Perspective scores alone; given that inter-rater agreement on these judgments is low (Krippendorff's $\alpha$ = 0.14), this performance approaches the ceiling that human disagreement imposes on individual-level prediction. At the group level, the model reproduces observed group removal rates within approximately one percentage point.

\textbf{Group-Specific Calibration on Perceived Toxicity.} The model reproduces group average removal rates but compresses the group-specific relationship between perceived toxicity and removal: on the most toxic comments, imputed between-group gaps are roughly half their observed size. This compression reflects an information limit: across six alternative architectures (including joint, sequential, and group-slope specifications), none recovered the relationship, because an unobserved respondent's perceived toxicity can be predicted only to within approximately 0.6 points on the 0-4 scale, which is a bound consistent with the low inter-rater agreement in these data. Because the simulation evaluates each group's outcomes against that group's own toxicity ratings, the relevant quantity is the relationship between removal and group-specific perceived toxicity. We therefore apply a group-specific calibration: for each group, we fit the observed removal rate as a smooth, monotonic function of the group's own mean perceived toxicity on the log-odds scale, and shift the imputed removal probabilities of that group's members to match the observed curve. The calibration sets group-level removal rates conditional on toxicity; the model continues to determine which comments within a toxicity level are more likely to be removed. At the respondent level, the corrections corresponding to each respondent's gender, racial, and LGB group memberships are applied jointly, and group marginal removal rates are preserved (within 0.011 overall, and within 0.003 for Black, Asian, Hispanic, and LGB respondents).

\textbf{Validation.} Figure S4 compares observed and imputed removal-rate gaps across comments binned by each focal group's own toxicity ratings. The imputed decision recovers the direction of the observed group-specific toxicity-removal relationship in all comparisons and 69-81\% of the observed gap magnitude in the highest-toxicity bin. The remaining difference reflects within-comment individual variation: in the observed data, each decision is coupled to the same respondent's perception of that comment, and a group-level calibration can restore this coupling only at the level of group rates, not for specific individuals.

\textbf{Use in the Simulation}. The imputed probabilities are used as expected values. For each comment and assigned panel, we compute the probability that the decision rule removes the comment (e.g., the probability that at least three of five members remove it) from the panel members' individual probabilities, and post-moderation toxicity weights each comment by its probability of remaining visible. Group toxicity outcomes are computed only on comments with at least one observed toxicity rating from that group (coverage in Table S9).

\newpage
\singlespacing
\section*{SI Figures}

\setcounter{figure}{0}\renewcommand{\thefigure}{S\arabic{figure}}

\begin{figure}[h]
\centering
\includegraphics[width=0.96\textwidth,height=0.82\textheight,keepaspectratio]{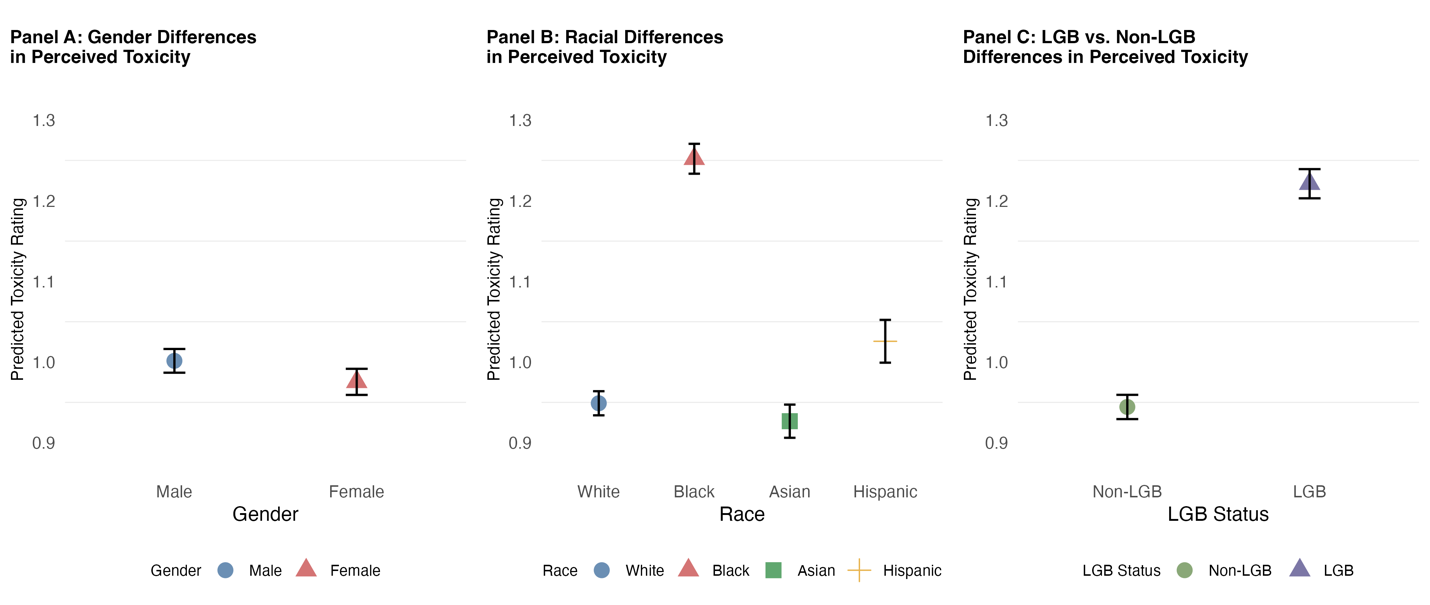}
\caption{\textbf{Group Differences in Predicted Toxicity Rating.} The plot shows the predicted toxicity rating of identical content by gender, race and LGB status. Error bars represent 95\% confidence intervals. Results are based on estimates from Model 2, SI Table S2.}
\label{fig:s1}
\end{figure}

\begin{figure}[p]
\centering
\includegraphics[width=0.96\textwidth,height=0.82\textheight,keepaspectratio]{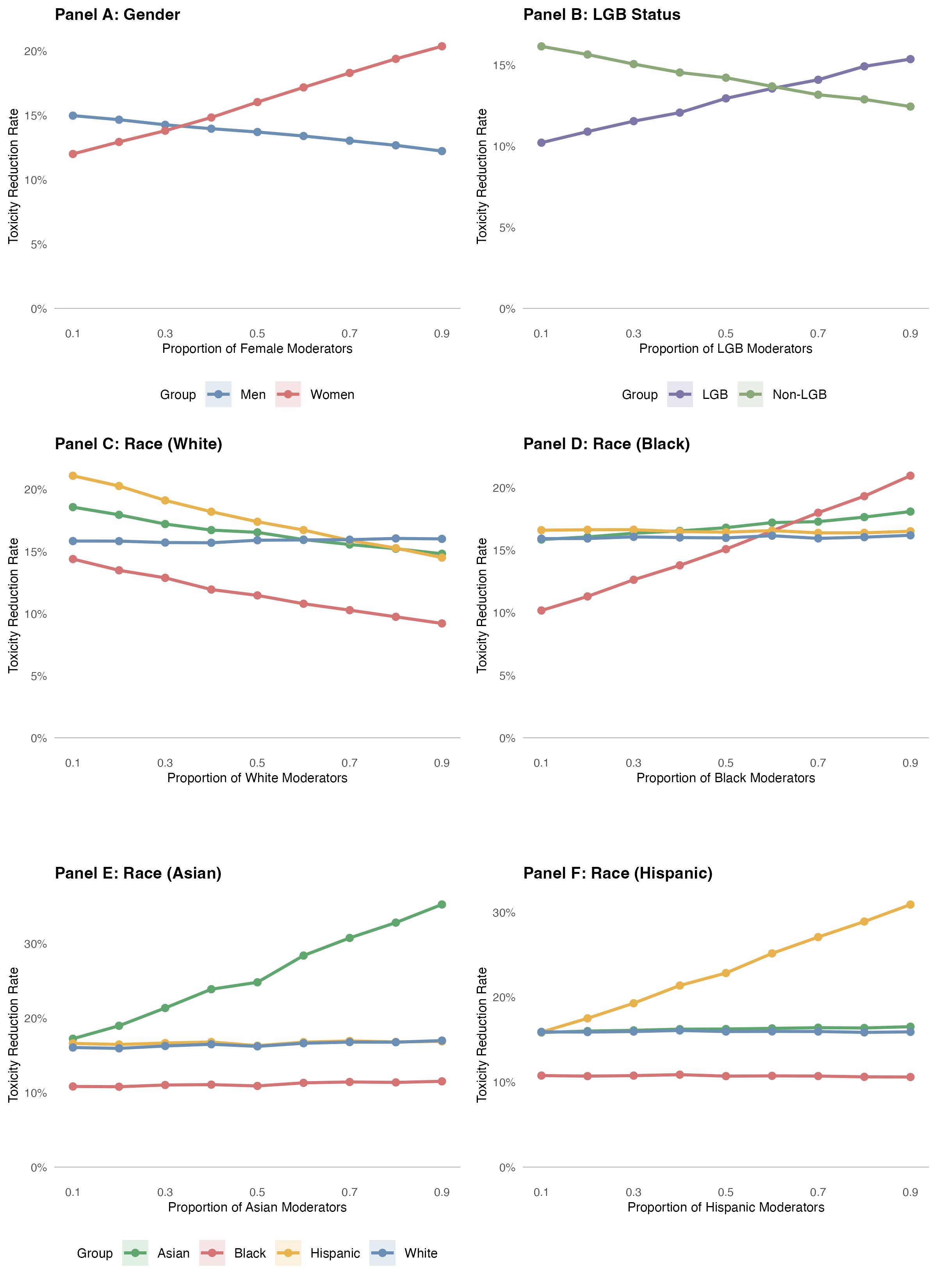}
\caption{\textbf{Simulated Relative Toxicity Reduction Across Moderator-Pool Compositions under the Majority-of-Five Rule.} This figure reproduces Figure 2 using the imputed data. The simulation is run on all 102,463 comments and applies a majority-of-five rule, under which a comment is removed when at least three of five assigned moderators would remove it. Error bands represent 95\% confidence intervals for the mean simulation estimate across 2,000 runs per scenario.}
\label{fig:s2}
\end{figure}

\begin{figure}[p]
\centering
\includegraphics[width=0.96\textwidth,height=0.82\textheight,keepaspectratio]{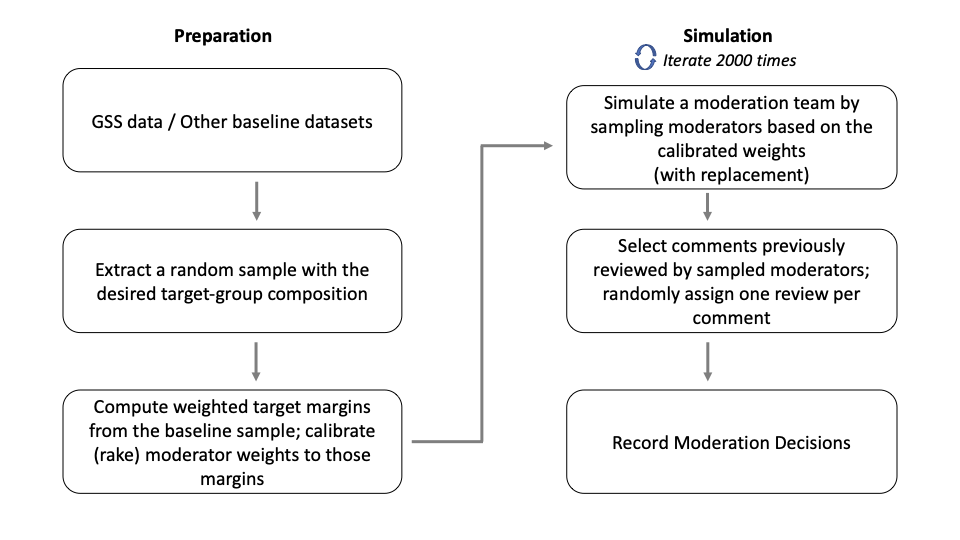}
\caption{\textbf{Simulation Procedure.} This figure shows the general workflow for both empirical-benchmark and counterfactual simulations. In counterfactual scenarios, we first construct modified target margins by resampling the baseline data to enforce a specified target-group proportion (e.g., 90\%). In empirical-benchmark simulations, this resampling step is skipped; target margins are computed directly from the observed survey-weighted distribution. In all cases, moderator weights are generated (via raking) to match the target margins before sampling moderator teams and recording moderation decisions. Moderators are sampled with replacement to allow stable draws for groups with limited representation in the data (e.g., Black, Asian, or LGB users).}
\label{fig:s3}
\end{figure}

\begin{figure}[p]
\centering
\includegraphics[width=0.96\textwidth,height=0.82\textheight,keepaspectratio]{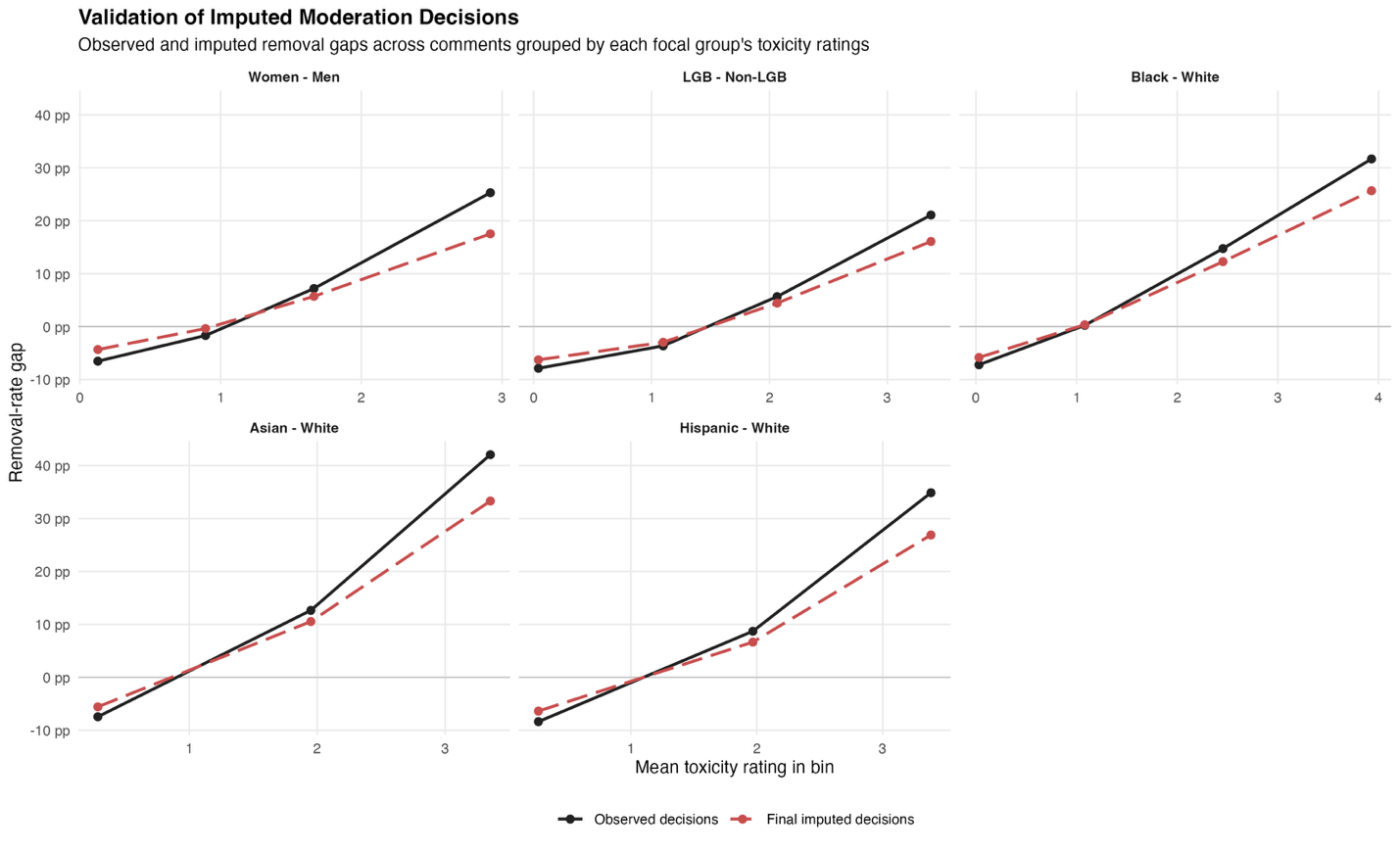}
\caption{\textbf{Validation of Imputed Moderation Decisions.} The figure shows observed and imputed removal-rate gaps across comments grouped by each focal group's own toxicity ratings. Each panels compares first-listed group to the second-listed group; the y-axis reports the first-listed group's removal rate minus the comparison group's removal rate. The black line uses observed moderation decisions. The red dashed line uses the imputed removal probabilities used in the alternative decision-rule simulations. The diagnostic shows that the imputed decisions preserve the observed group-specific relationship between perceived toxicity and removal: as comments become more toxic to the focal group, the imputed removal gaps move in the same direction and recovers most of the observed gap. In the highest-toxicity bin, the imputed gaps recover 69\% to 81\% of the observed gaps across the five panels. The remaining difference is expected, because, in the observed data, each comment's group-specific toxicity rating is tied to moderation decisions made by respondents who actually rated that comment. The imputed data extends moderation decisions to respondents who did not rate the comment, so it can restore the group-level relationship between perceived toxicity and removal where the data support it, but cannot fully recover within-comment individual variation in how specific respondents perceived and judged the same comment.}
\label{fig:s4}
\end{figure}

\begin{figure}[p]
\centering
\includegraphics[width=0.96\textwidth,height=0.82\textheight,keepaspectratio]{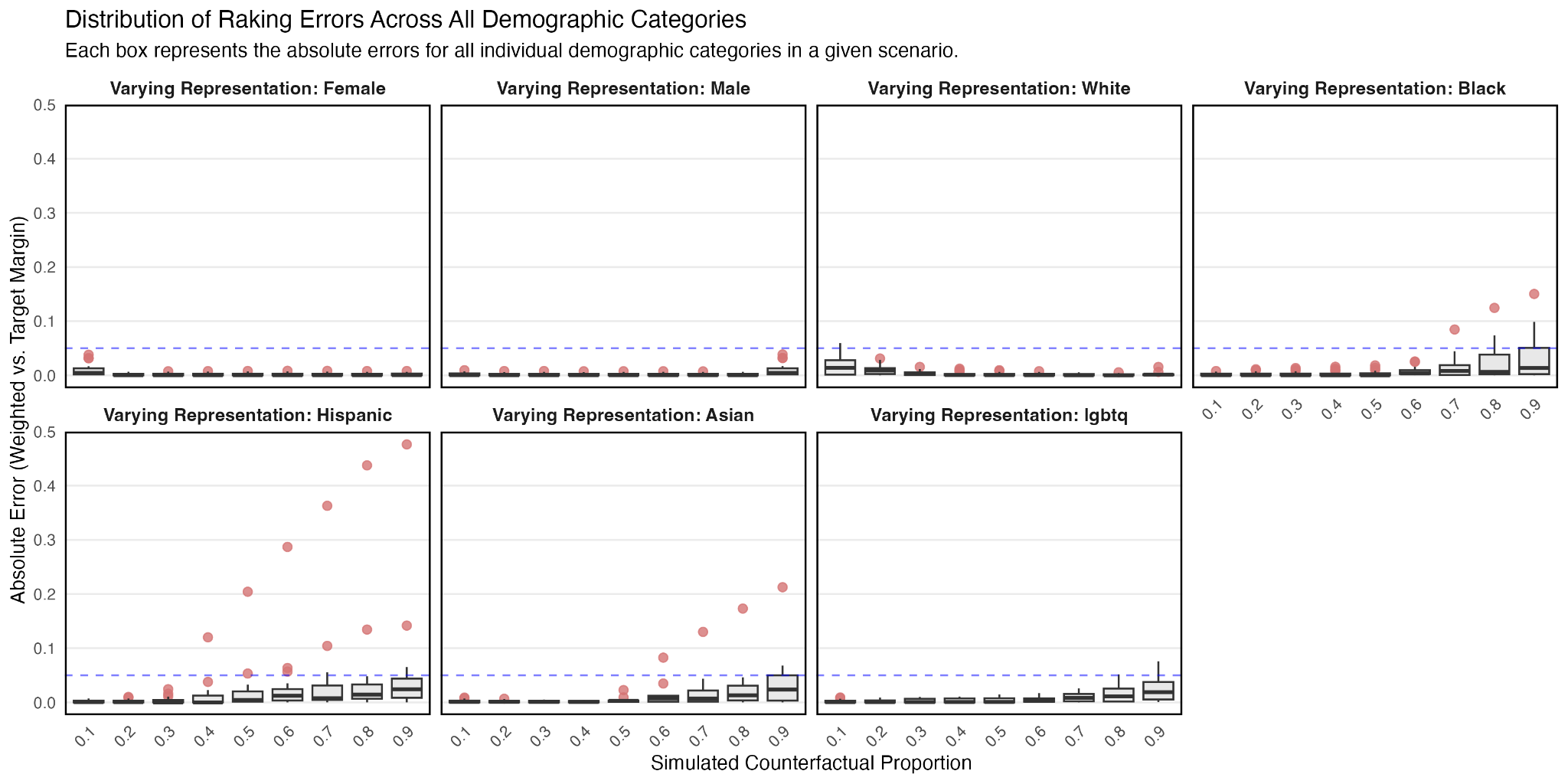}
\caption{\textbf{Distribution of Raking Convergence Errors in Observed-Data Simulations (Counterfactual Simulation Scenarios).} To verify the structural validity of counterfactual moderator weights, we calculated the absolute error between the target baseline margins and the achieved post-raking margins for all 20 individual demographic sub-categories (spanning age, race, gender, education, and LGB status). This figure plots the distribution of errors for each simulated target proportion, expressed as decimal proportions, where a value of 0.05 represents a 5-percentage-point deviation. The boxplots demonstrate that median absolute error remains at or near zero across almost all simulations, indicating successful multidimensional balancing. For extreme counterfactual scenarios (e.g., simulating a 90\% Hispanic moderator team), generating perfectly balanced multidimensional weights is constrained by the inherent demographic distributions within the underlying survey data (e.g., the scarcity of older Hispanic respondents). To prevent a small number of intersecting respondents from receiving disproportionately massive weights and artificially inflating the variance of the simulation, we capped the maximum raking weight at 10. The isolated red outliers at extreme target boundaries represent the algorithm caps weights to prevent heavily up-weighting sparse intersecting cells (such as older, minority respondents), thereby prioritizing simulation variance control over perfect convergence in extreme counterfactuals.}
\label{fig:s5}
\end{figure}

\begin{figure}[p]
\centering
\includegraphics[width=0.96\textwidth,height=0.82\textheight,keepaspectratio]{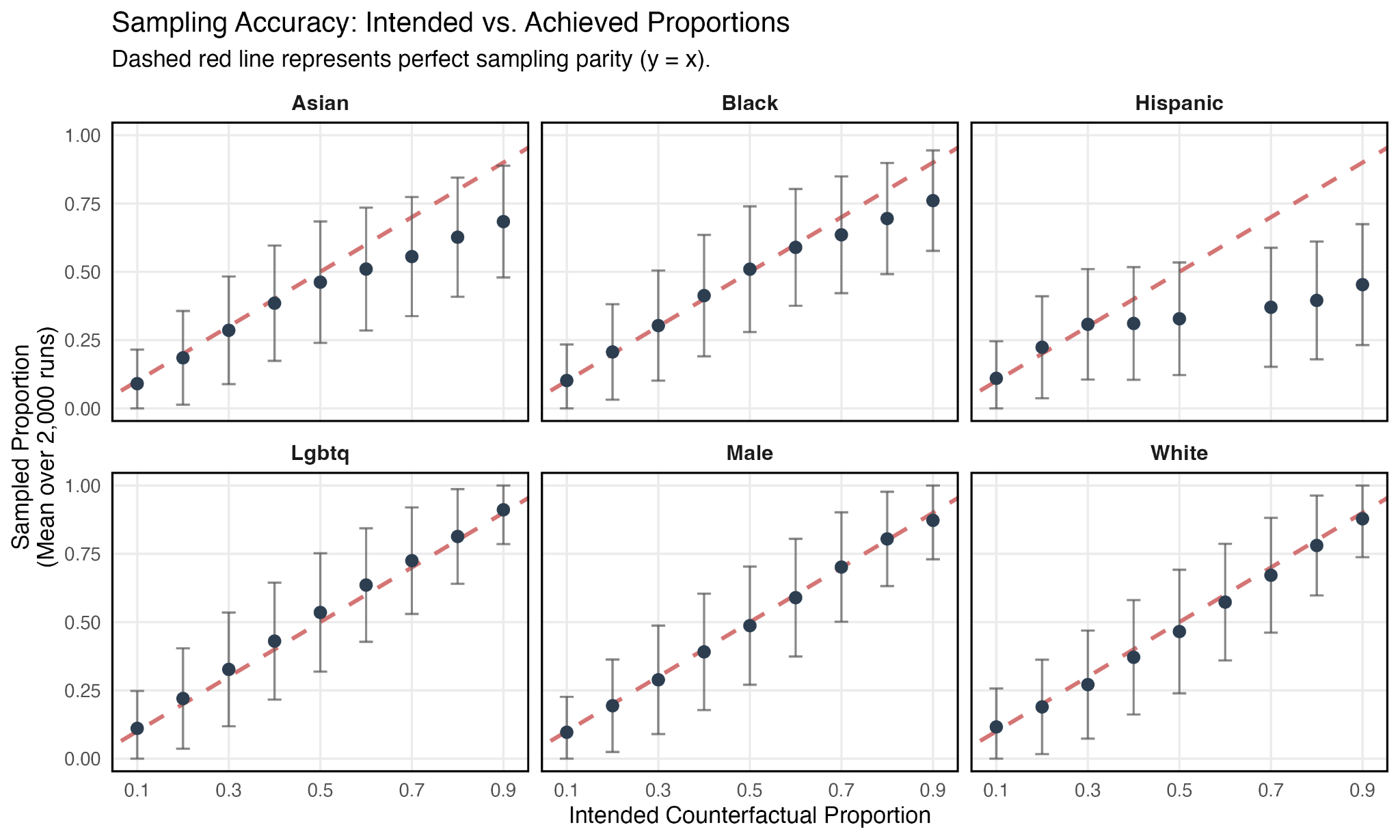}
\caption{\textbf{Sampling Accuracy for Observed-Data Simulations (Counterfactual Simulation Scenarios).} This plot shows the accuracy of the sampling mechanism across all counterfactual scenarios. The x-axis denotes the intended counterfactual proportion defined in the simulation design, while the y-axis denotes the empirical mean of the actual proportion sampled across 2,000 iterations. Data points aligning with the 45-degree parity line (dashed red) indicating that the weighted sampling procedure faithfully executes the target design. The error bars represent the 95\% confidence interval of the sampled proportions across the 2,000 runs. The sampled proportion flattens out for some minority groups at high targets (such as \textgreater{} 60\% Hispanic moderators). This is the downstream result of the weight cap applied during the raking phase.}
\label{fig:s6}
\end{figure}

\begin{figure}[p]
\centering
\includegraphics[width=0.96\textwidth,height=0.82\textheight,keepaspectratio]{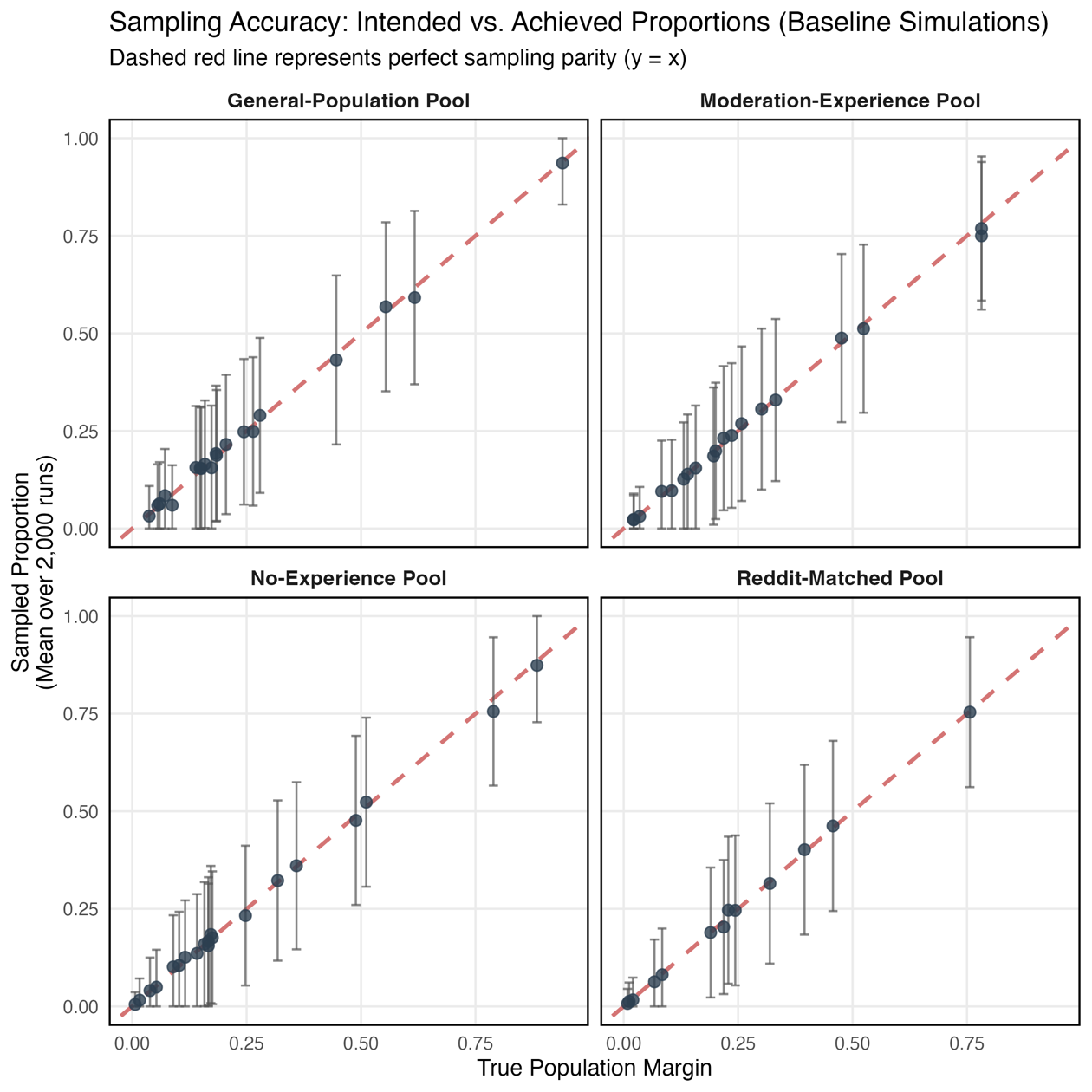}
\caption{\textbf{Sampling Accuracy for Observed-Data Simulations (Benchmark Simulation Scenarios).} The x-axis denotes the true empirical proportion of each demographic sub-category (spanning age, race, gender, education, and LGB status) within the respective reference population. The y-axis denotes the empirical mean of the actual proportion sampled for that specific sub-category across the 2,000 simulation iterations. Each point represents a distinct demographic category. The error bars represent the 95\% confidence interval of the sampled proportions across the 2,000 runs.}
\label{fig:s7}
\end{figure}

\begin{figure}[p]
\centering
\includegraphics[width=0.96\textwidth,height=0.82\textheight,keepaspectratio]{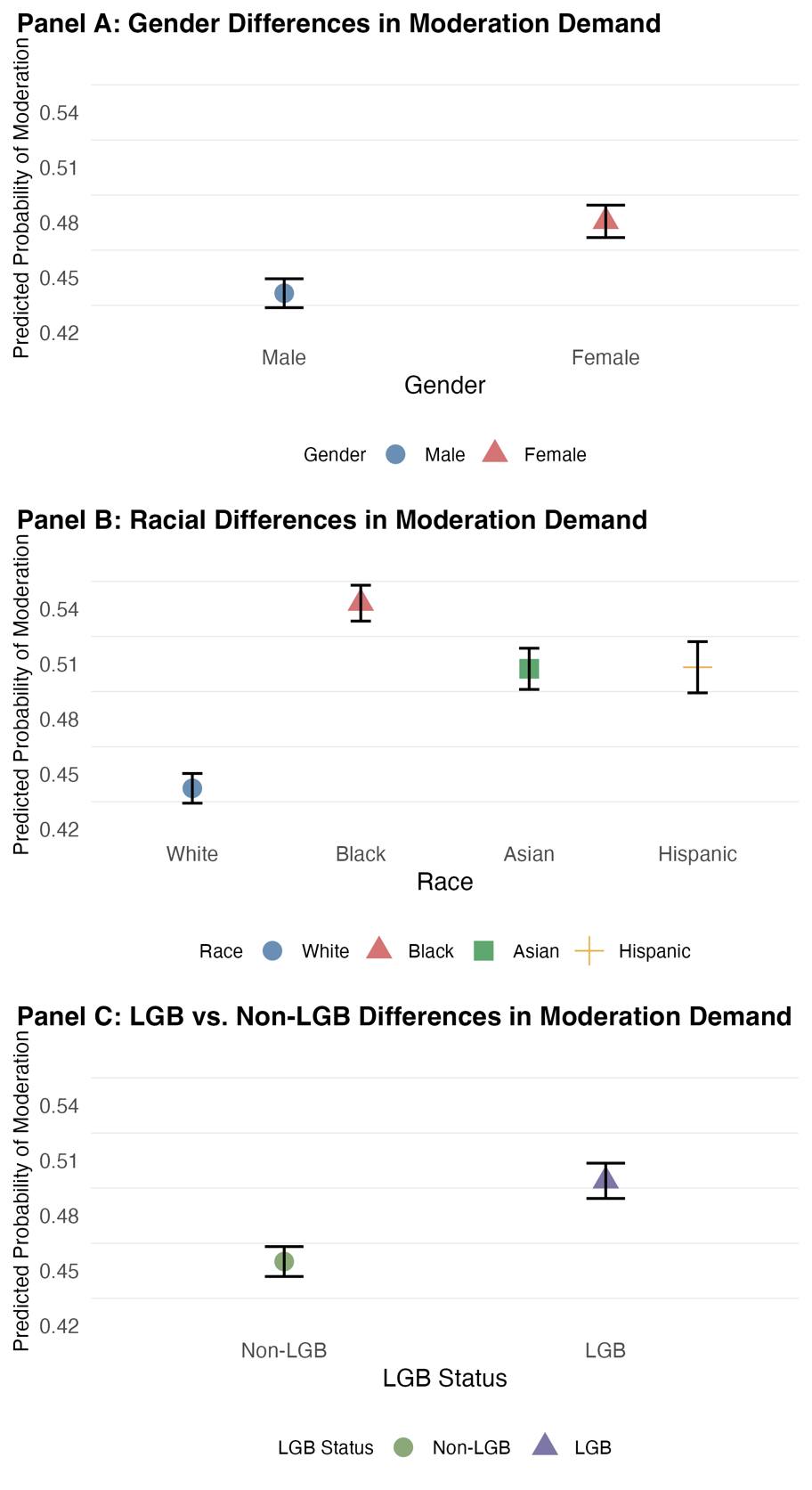}
\caption{\textbf{Group Differences in Predicted Probability of Moderation Demand When ``It Depends on the Context'' Coded as Removal.} The plot shows the predicted moderation demand by gender, race and LGB status. Error bars represent 95\% confidence intervals.}
\label{fig:s8}
\end{figure}

\begin{figure}[p]
\centering
\includegraphics[width=0.96\textwidth,height=0.82\textheight,keepaspectratio]{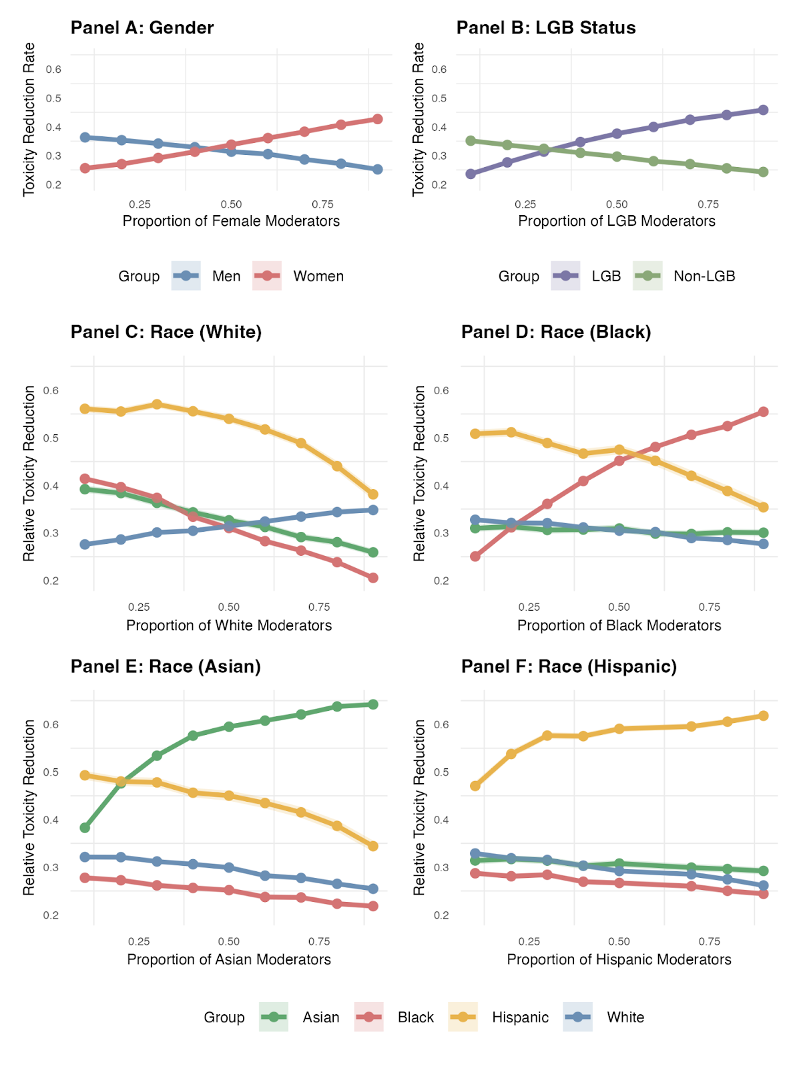}
\caption{\textbf{Simulated Effects of Varying Moderator Pool Composition on Relative Toxicity Reduction Across Groups When ``It Depends on the Context'' Coded as Moderated.} Error bands represent 95\% confidence intervals for the mean simulation estimate across 2,000 runs per scenario.}
\label{fig:s9}
\end{figure}

\begin{figure}[p]
\centering
\includegraphics[width=0.96\textwidth,height=0.82\textheight,keepaspectratio]{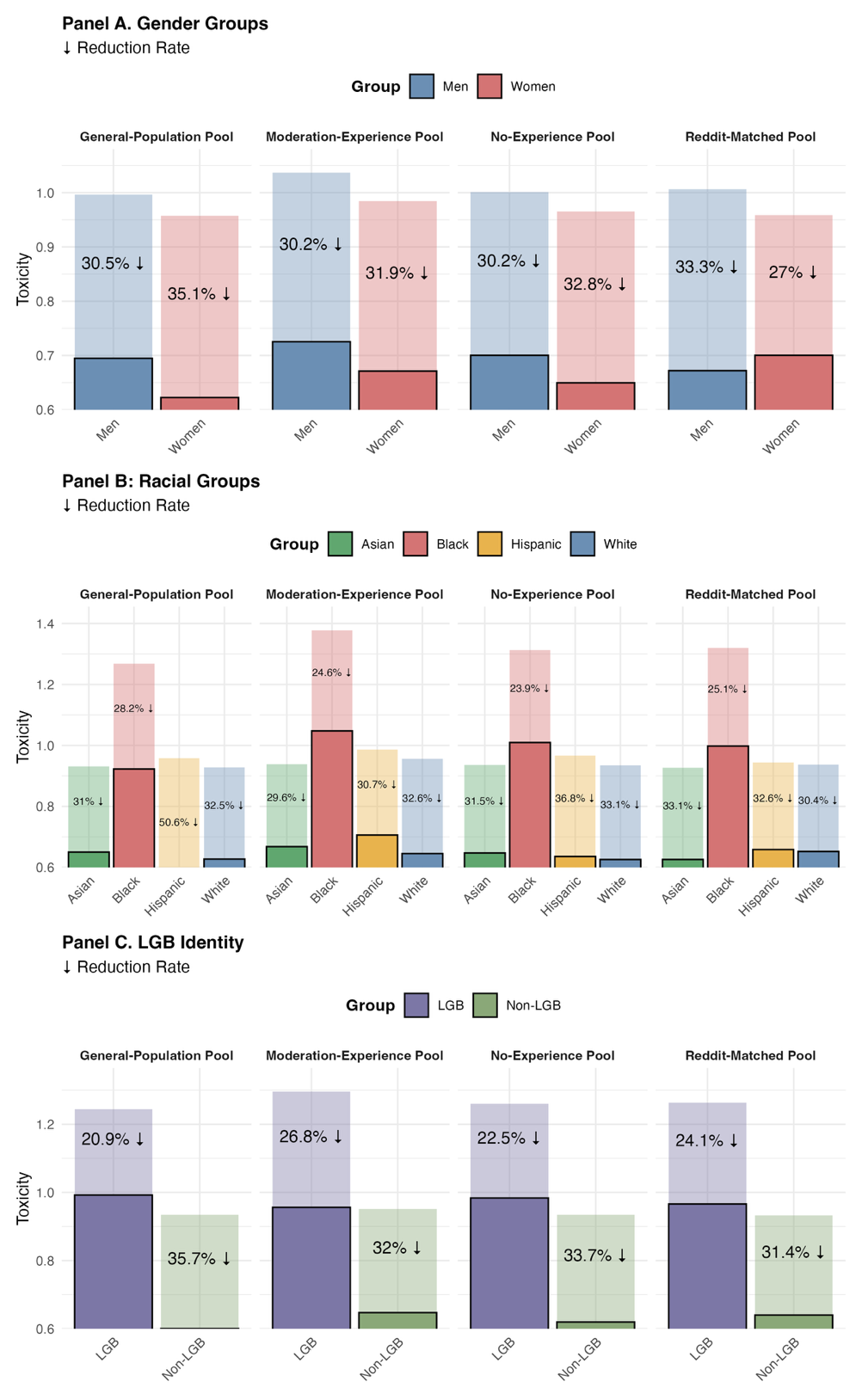}
\caption{\textbf{Simulated Relative Toxicity Reduction Across Four Empirically Grounded Moderation Scenarios (General Population, Moderation-Experience, No-Experience, and Reddit-Matched Pools) When ``It Depends on the Context'' Coded as Moderated.} Shaded background bars show mean pre-moderation toxicity. Solid bars show post-moderation toxicity. Annotations denote relative toxicity reduction. Values are means across 2000 runs per scenario.}
\label{fig:s10}
\end{figure}

\begin{figure}[p]
\centering
\includegraphics[width=0.96\textwidth,height=0.82\textheight,keepaspectratio]{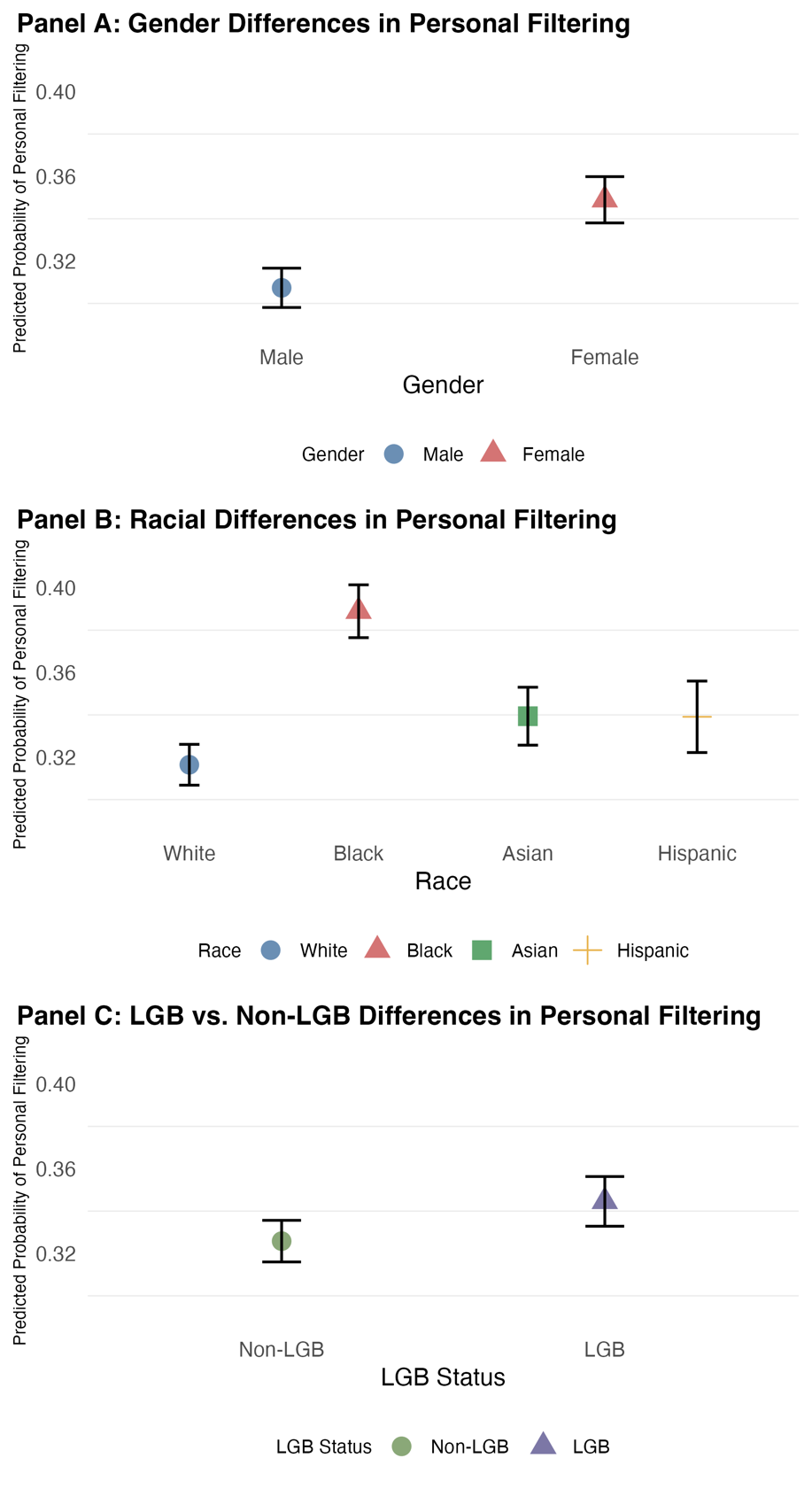}
\caption{\textbf{Group Differences in Predicted Probability of Moderation Demand (Moderation Based on Personal Filtering Preference: ``It's Fine for Me to See'' vs. ``I Would Never Want to See It Online'').} The plot shows the predicted moderation demand by gender, race and LGB status. Error bars represent 95\% confidence intervals.}
\label{fig:s11}
\end{figure}

\begin{figure}[p]
\centering
\includegraphics[width=0.96\textwidth,height=0.82\textheight,keepaspectratio]{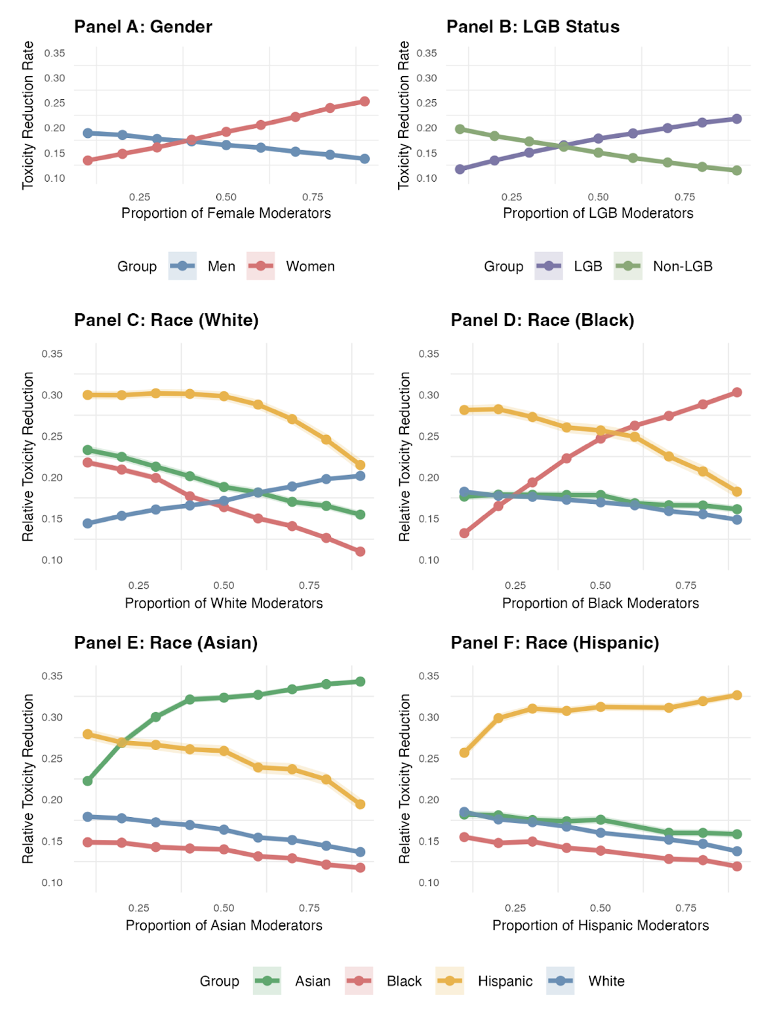}
\caption{\textbf{Simulated Effects of Varying Moderator Pool Composition on Relative Toxicity Reduction Across Groups (Moderation Based on Personal Filtering Preference: ``It's Fine for Me to See'' vs. ``I Would Never Want to See It Online'').} Error bands represent 95\% confidence intervals for the mean simulation estimate across 2,000 runs per scenario.}
\label{fig:s12}
\end{figure}

\begin{figure}[p]
\centering
\includegraphics[width=0.96\textwidth,height=0.82\textheight,keepaspectratio]{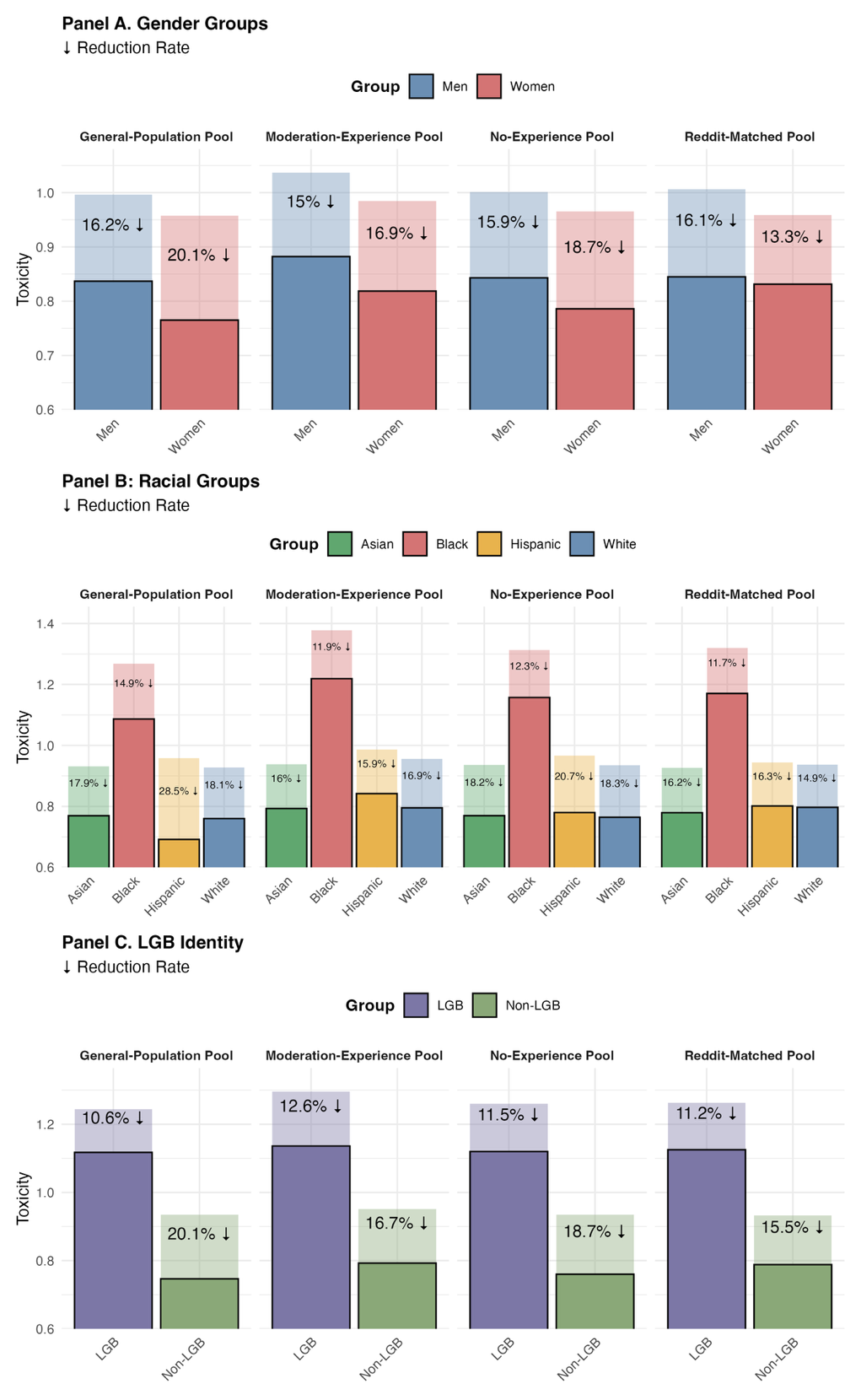}
\caption{\textbf{Simulated Relative Toxicity Reduction Across Four Empirically Grounded Moderation Scenarios (General-Population, Moderation-Experience, No-Experience, and Reddit-Matched Pools) When Moderation Decisions are Based on Personal Filtering Preference: ``It's Fine for Me to See'' vs. ``I Would Never Want to See It Online'').} Solid bars show post-moderation toxicity. Annotations denote relative toxicity reduction. Values are means across 2000 runs per scenario.}
\label{fig:s13}
\end{figure}

\clearpage\setcounter{table}{0}\renewcommand{\thetable}{S\arabic{table}}

\section*{SI Tables}
\noindent\textbf{Table S1. Descriptive Statistics of Toxicity Ratings and Moderation Decisions}\par\smallskip

\begin{longtable}[]{@{}
  >{\raggedright\arraybackslash}p{(\linewidth - 2\tabcolsep) * \real{0.3892}}
  >{\centering\arraybackslash}p{(\linewidth - 2\tabcolsep) * \real{0.6108}}@{}}
\toprule\noalign{}
\endhead
\bottomrule\noalign{}
\endlastfoot
& Full Sample

(N = 501,540) \\
& Mean (SD) / \% \\
\emph{Perceived Toxicity (Continuous)} & 0.99

(1.27) \\
\emph{Perceived Toxicity (Categorical)} & \\
Not at all toxic & 52.33 \\
Slightly toxic & 18.56 \\
Moderately toxic & 13.27 \\
Very toxic & 9.37 \\
Extremely toxic & 6.47 \\
\emph{Moderation Decision (Binary)} & \\
Not moderated & 84.61 \\
Moderated & 15.39 \\
\emph{Moderation Decision (Original Categories)} & \\
This comment should be allowed & 57.33 \\
It depends on the context & 27.28 \\
This comment should be removed & 15.39 \\
\end{longtable}

\textbf{Notes:} (1) The unit of observation is an individual respondent--comment rating. The analytical sample comprises 501,540 ratings from 16,221 respondents evaluating 102,463 comments. (2) Moderation Decision (Binary) is recoded based on Moderation Decision (Original Categories): ``This comment should be allowed'' or ``It depends on the context'' = Not moderated; ``This comment should be removed'' = Moderated. The unit of observation is an individual respondent--comment rating.

\clearpage
\noindent\textbf{Table S2. Regression Estimates for Moderation Demand and Toxicity Perception with Comment Fixed Effects.}\par\smallskip

\begin{longtable}[]{@{}
  >{\raggedright\arraybackslash}p{(\linewidth - 10\tabcolsep) * \real{0.2729}}
  >{\raggedright\arraybackslash}p{(\linewidth - 10\tabcolsep) * \real{0.1454}}
  >{\raggedright\arraybackslash}p{(\linewidth - 10\tabcolsep) * \real{0.1503}}
  >{\raggedright\arraybackslash}p{(\linewidth - 10\tabcolsep) * \real{0.1405}}
  >{\raggedright\arraybackslash}p{(\linewidth - 10\tabcolsep) * \real{0.1454}}
  >{\raggedright\arraybackslash}p{(\linewidth - 10\tabcolsep) * \real{0.1454}}@{}}
\toprule\noalign{}
\endhead
\bottomrule\noalign{}
\endlastfoot
& \begin{minipage}[t]{\linewidth}\raggedright

Model 1

\end{minipage} & \begin{minipage}[t]{\linewidth}\raggedright

Model 2

\end{minipage} & \begin{minipage}[t]{\linewidth}\raggedright

Model 3

\end{minipage} & \begin{minipage}[t]{\linewidth}\raggedright

Model 4

\end{minipage} & \begin{minipage}[t]{\linewidth}\raggedright

Model 5

\end{minipage} \\
\begin{minipage}[t]{\linewidth}\raggedright

Dependent Variable

\end{minipage} & \begin{minipage}[t]{\linewidth}\raggedright

Moderation Demand

\end{minipage} & \begin{minipage}[t]{\linewidth}\raggedright

Toxicity Perception

\end{minipage} & \begin{minipage}[t]{\linewidth}\raggedright

Moderation Demand

\end{minipage} & \begin{minipage}[t]{\linewidth}\raggedright

Moderation Demand

\end{minipage} & \begin{minipage}[t]{\linewidth}\raggedright

Moderation Demand

\end{minipage} \\
\begin{minipage}[t]{\linewidth}\raggedright

Female

\end{minipage} & \begin{minipage}[t]{\linewidth}\raggedright

0.176***

\end{minipage} & \begin{minipage}[t]{\linewidth}\raggedright

-0.026***

\end{minipage} & \begin{minipage}[t]{\linewidth}\raggedright

-0.099**

\end{minipage} & \begin{minipage}[t]{\linewidth}\raggedright

0.284***

\end{minipage} & \begin{minipage}[t]{\linewidth}\raggedright

0.284***

\end{minipage} \\
& \begin{minipage}[t]{\linewidth}\raggedright

(0.012)

\end{minipage} & \begin{minipage}[t]{\linewidth}\raggedright

(0.003)

\end{minipage} & \begin{minipage}[t]{\linewidth}\raggedright

(0.029)

\end{minipage} & \begin{minipage}[t]{\linewidth}\raggedright

(0.018)

\end{minipage} & \begin{minipage}[t]{\linewidth}\raggedright

(0.018)

\end{minipage} \\
\begin{minipage}[t]{\linewidth}\raggedright

Race (Ref. = White)

\end{minipage} & & & & & \\
\begin{minipage}[t]{\linewidth}\raggedright

Asian

\end{minipage} & \begin{minipage}[t]{\linewidth}\raggedright

0.289***

\end{minipage} & \begin{minipage}[t]{\linewidth}\raggedright

-0.022***

\end{minipage} & \begin{minipage}[t]{\linewidth}\raggedright

0.49***

\end{minipage} & \begin{minipage}[t]{\linewidth}\raggedright

0.157*

\end{minipage} & \begin{minipage}[t]{\linewidth}\raggedright

0.505***

\end{minipage} \\
& \begin{minipage}[t]{\linewidth}\raggedright

(0.025)

\end{minipage} & \begin{minipage}[t]{\linewidth}\raggedright

(0.007)

\end{minipage} & \begin{minipage}[t]{\linewidth}\raggedright

(0.036)

\end{minipage} & \begin{minipage}[t]{\linewidth}\raggedright

(0.064)

\end{minipage} & \begin{minipage}[t]{\linewidth}\raggedright

(0.036)

\end{minipage} \\
\begin{minipage}[t]{\linewidth}\raggedright

Black

\end{minipage} & \begin{minipage}[t]{\linewidth}\raggedright

0.465***

\end{minipage} & \begin{minipage}[t]{\linewidth}\raggedright

0.303***

\end{minipage} & \begin{minipage}[t]{\linewidth}\raggedright

0.064*

\end{minipage} & \begin{minipage}[t]{\linewidth}\raggedright

1.10***

\end{minipage} & \begin{minipage}[t]{\linewidth}\raggedright

0.104***

\end{minipage} \\
& \begin{minipage}[t]{\linewidth}\raggedright

(0.018)

\end{minipage} & \begin{minipage}[t]{\linewidth}\raggedright

(0.006)

\end{minipage} & \begin{minipage}[t]{\linewidth}\raggedright

(0.028)

\end{minipage} & \begin{minipage}[t]{\linewidth}\raggedright

(0.041)

\end{minipage} & \begin{minipage}[t]{\linewidth}\raggedright

(0.028)

\end{minipage} \\
\begin{minipage}[t]{\linewidth}\raggedright

Hispanic

\end{minipage} & \begin{minipage}[t]{\linewidth}\raggedright

0.179***

\end{minipage} & \begin{minipage}[t]{\linewidth}\raggedright

0.077***

\end{minipage} & \begin{minipage}[t]{\linewidth}\raggedright

0.067

\end{minipage} & \begin{minipage}[t]{\linewidth}\raggedright

0.189*

\end{minipage} & \begin{minipage}[t]{\linewidth}\raggedright

0.086

\end{minipage} \\
& \begin{minipage}[t]{\linewidth}\raggedright

(0.037)

\end{minipage} & \begin{minipage}[t]{\linewidth}\raggedright

(0.011)

\end{minipage} & \begin{minipage}[t]{\linewidth}\raggedright

(0.056)

\end{minipage} & \begin{minipage}[t]{\linewidth}\raggedright

(0.088)

\end{minipage} & \begin{minipage}[t]{\linewidth}\raggedright

(0.056)

\end{minipage} \\
\begin{minipage}[t]{\linewidth}\raggedright

Multiracial

\end{minipage} & \begin{minipage}[t]{\linewidth}\raggedright

-0.076*

\end{minipage} & \begin{minipage}[t]{\linewidth}\raggedright

-0.062***

\end{minipage} & \begin{minipage}[t]{\linewidth}\raggedright

0.018

\end{minipage} & \begin{minipage}[t]{\linewidth}\raggedright

-0.34***

\end{minipage} & \begin{minipage}[t]{\linewidth}\raggedright

0.021

\end{minipage} \\
& \begin{minipage}[t]{\linewidth}\raggedright

(0.031)

\end{minipage} & \begin{minipage}[t]{\linewidth}\raggedright

(0.008)

\end{minipage} & \begin{minipage}[t]{\linewidth}\raggedright

(0.045)

\end{minipage} & \begin{minipage}[t]{\linewidth}\raggedright

(0.081)

\end{minipage} & \begin{minipage}[t]{\linewidth}\raggedright

(0.045)

\end{minipage} \\
\begin{minipage}[t]{\linewidth}\raggedright

Other

\end{minipage} & \begin{minipage}[t]{\linewidth}\raggedright

0.255***

\end{minipage} & \begin{minipage}[t]{\linewidth}\raggedright

0.160***

\end{minipage} & \begin{minipage}[t]{\linewidth}\raggedright

0.058

\end{minipage} & \begin{minipage}[t]{\linewidth}\raggedright

0.559***

\end{minipage} & \begin{minipage}[t]{\linewidth}\raggedright

0.047

\end{minipage} \\
& \begin{minipage}[t]{\linewidth}\raggedright

(0.046)

\end{minipage} & \begin{minipage}[t]{\linewidth}\raggedright

(0.014)

\end{minipage} & \begin{minipage}[t]{\linewidth}\raggedright

(0.070)

\end{minipage} & \begin{minipage}[t]{\linewidth}\raggedright

(0.108)

\end{minipage} & \begin{minipage}[t]{\linewidth}\raggedright

(0.070)

\end{minipage} \\
\begin{minipage}[t]{\linewidth}\raggedright

Education

(Ref. = High Schools)

\end{minipage} & & & & & \\
\begin{minipage}[t]{\linewidth}\raggedright

College

\end{minipage} & \begin{minipage}[t]{\linewidth}\raggedright

0.055**

\end{minipage} & \begin{minipage}[t]{\linewidth}\raggedright

0.149***

\end{minipage} & \begin{minipage}[t]{\linewidth}\raggedright

-0.165***

\end{minipage} & \begin{minipage}[t]{\linewidth}\raggedright

-0.164***

\end{minipage} & \begin{minipage}[t]{\linewidth}\raggedright

-0.163***

\end{minipage} \\
& \begin{minipage}[t]{\linewidth}\raggedright

(0.022)

\end{minipage} & \begin{minipage}[t]{\linewidth}\raggedright

(0.006)

\end{minipage} & \begin{minipage}[t]{\linewidth}\raggedright

(0.032)

\end{minipage} & \begin{minipage}[t]{\linewidth}\raggedright

(0.033)

\end{minipage} & \begin{minipage}[t]{\linewidth}\raggedright

(0.033)

\end{minipage} \\
\begin{minipage}[t]{\linewidth}\raggedright

Graduate

\end{minipage} & \begin{minipage}[t]{\linewidth}\raggedright

-0.029

\end{minipage} & \begin{minipage}[t]{\linewidth}\raggedright

0.209***

\end{minipage} & \begin{minipage}[t]{\linewidth}\raggedright

-0.466***

\end{minipage} & \begin{minipage}[t]{\linewidth}\raggedright

-0.436***

\end{minipage} & \begin{minipage}[t]{\linewidth}\raggedright

-0.458***

\end{minipage} \\
& \begin{minipage}[t]{\linewidth}\raggedright

(0.024)

\end{minipage} & \begin{minipage}[t]{\linewidth}\raggedright

(0.007)

\end{minipage} & \begin{minipage}[t]{\linewidth}\raggedright

(0.036)

\end{minipage} & \begin{minipage}[t]{\linewidth}\raggedright

(0.037)

\end{minipage} & \begin{minipage}[t]{\linewidth}\raggedright

(0.037)

\end{minipage} \\
\begin{minipage}[t]{\linewidth}\raggedright

Less than High School

\end{minipage} & \begin{minipage}[t]{\linewidth}\raggedright

0.288***

\end{minipage} & \begin{minipage}[t]{\linewidth}\raggedright

0.108***

\end{minipage} & \begin{minipage}[t]{\linewidth}\raggedright

0.208

\end{minipage} & \begin{minipage}[t]{\linewidth}\raggedright

0.217

\end{minipage} & \begin{minipage}[t]{\linewidth}\raggedright

0.230

\end{minipage} \\
& \begin{minipage}[t]{\linewidth}\raggedright

(0.082)

\end{minipage} & \begin{minipage}[t]{\linewidth}\raggedright

(0.022)

\end{minipage} & \begin{minipage}[t]{\linewidth}\raggedright

(0.116)

\end{minipage} & \begin{minipage}[t]{\linewidth}\raggedright

(0.117)

\end{minipage} & \begin{minipage}[t]{\linewidth}\raggedright

(0.116)

\end{minipage} \\
\begin{minipage}[t]{\linewidth}\raggedright

Some College

\end{minipage} & \begin{minipage}[t]{\linewidth}\raggedright

-0.147***

\end{minipage} & \begin{minipage}[t]{\linewidth}\raggedright

-0.038***

\end{minipage} & \begin{minipage}[t]{\linewidth}\raggedright

-0.113***

\end{minipage} & \begin{minipage}[t]{\linewidth}\raggedright

-0.112***

\end{minipage} & \begin{minipage}[t]{\linewidth}\raggedright

-0.120***

\end{minipage} \\
& \begin{minipage}[t]{\linewidth}\raggedright

(0.022)

\end{minipage} & \begin{minipage}[t]{\linewidth}\raggedright

(0.006)

\end{minipage} & \begin{minipage}[t]{\linewidth}\raggedright

(0.033)

\end{minipage} & \begin{minipage}[t]{\linewidth}\raggedright

(0.033)

\end{minipage} & \begin{minipage}[t]{\linewidth}\raggedright

(0.033)

\end{minipage} \\
\begin{minipage}[t]{\linewidth}\raggedright

Age (Ref. = 18-24)

\end{minipage} & & & & & \\
25-34 & \begin{minipage}[t]{\linewidth}\raggedright

0.071***

\end{minipage} & \begin{minipage}[t]{\linewidth}\raggedright

0.048***

\end{minipage} & \begin{minipage}[t]{\linewidth}\raggedright

0.069*

\end{minipage} & \begin{minipage}[t]{\linewidth}\raggedright

0.075*

\end{minipage} & \begin{minipage}[t]{\linewidth}\raggedright

0.086*

\end{minipage} \\
& \begin{minipage}[t]{\linewidth}\raggedright

(0.021)

\end{minipage} & \begin{minipage}[t]{\linewidth}\raggedright

(0.006)

\end{minipage} & \begin{minipage}[t]{\linewidth}\raggedright

(0.031)

\end{minipage} & \begin{minipage}[t]{\linewidth}\raggedright

(0.031)

\end{minipage} & \begin{minipage}[t]{\linewidth}\raggedright

(0.031)

\end{minipage} \\
\begin{minipage}[t]{\linewidth}\raggedright

35-44

\end{minipage} & \begin{minipage}[t]{\linewidth}\raggedright

-0.021

\end{minipage} & \begin{minipage}[t]{\linewidth}\raggedright

-0.030***

\end{minipage} & \begin{minipage}[t]{\linewidth}\raggedright

0.087*

\end{minipage} & \begin{minipage}[t]{\linewidth}\raggedright

0.098**

\end{minipage} & \begin{minipage}[t]{\linewidth}\raggedright

0.115***

\end{minipage} \\
& \begin{minipage}[t]{\linewidth}\raggedright

(0.022)

\end{minipage} & \begin{minipage}[t]{\linewidth}\raggedright

(0.006)

\end{minipage} & \begin{minipage}[t]{\linewidth}\raggedright

(0.033)

\end{minipage} & \begin{minipage}[t]{\linewidth}\raggedright

(0.033)

\end{minipage} & \begin{minipage}[t]{\linewidth}\raggedright

(0.033)

\end{minipage} \\
\begin{minipage}[t]{\linewidth}\raggedright

45-54

\end{minipage} & \begin{minipage}[t]{\linewidth}\raggedright

0.096***

\end{minipage} & \begin{minipage}[t]{\linewidth}\raggedright

-0.034*

\end{minipage} & \begin{minipage}[t]{\linewidth}\raggedright

0.263***

\end{minipage} & \begin{minipage}[t]{\linewidth}\raggedright

0.278***

\end{minipage} & \begin{minipage}[t]{\linewidth}\raggedright

0.301***

\end{minipage} \\
& \begin{minipage}[t]{\linewidth}\raggedright

(0.025)

\end{minipage} & \begin{minipage}[t]{\linewidth}\raggedright

(0.007)

\end{minipage} & \begin{minipage}[t]{\linewidth}\raggedright

(0.037)

\end{minipage} & \begin{minipage}[t]{\linewidth}\raggedright

(0.037)

\end{minipage} & \begin{minipage}[t]{\linewidth}\raggedright

(0.037)

\end{minipage} \\
\begin{minipage}[t]{\linewidth}\raggedright

55-64

\end{minipage} & \begin{minipage}[t]{\linewidth}\raggedright

0.251***

\end{minipage} & \begin{minipage}[t]{\linewidth}\raggedright

-0.004

\end{minipage} & \begin{minipage}[t]{\linewidth}\raggedright

0.441***

\end{minipage} & \begin{minipage}[t]{\linewidth}\raggedright

0.467***

\end{minipage} & \begin{minipage}[t]{\linewidth}\raggedright

0.473***

\end{minipage} \\
& \begin{minipage}[t]{\linewidth}\raggedright

(0.029)

\end{minipage} & \begin{minipage}[t]{\linewidth}\raggedright

(0.008)

\end{minipage} & \begin{minipage}[t]{\linewidth}\raggedright

(0.043)

\end{minipage} & \begin{minipage}[t]{\linewidth}\raggedright

(0.043)

\end{minipage} & \begin{minipage}[t]{\linewidth}\raggedright

(0.043)

\end{minipage} \\
\begin{minipage}[t]{\linewidth}\raggedright

65 or older

\end{minipage} & \begin{minipage}[t]{\linewidth}\raggedright

0.510***

\end{minipage} & \begin{minipage}[t]{\linewidth}\raggedright

0.066***

\end{minipage} & \begin{minipage}[t]{\linewidth}\raggedright

0.593***

\end{minipage} & \begin{minipage}[t]{\linewidth}\raggedright

0.614***

\end{minipage} & \begin{minipage}[t]{\linewidth}\raggedright

0.631***

\end{minipage} \\
& \begin{minipage}[t]{\linewidth}\raggedright

(0.038)

\end{minipage} & \begin{minipage}[t]{\linewidth}\raggedright

(0.011)

\end{minipage} & \begin{minipage}[t]{\linewidth}\raggedright

(0.056)

\end{minipage} & \begin{minipage}[t]{\linewidth}\raggedright

(0.056)

\end{minipage} & \begin{minipage}[t]{\linewidth}\raggedright

(0.056)

\end{minipage} \\
\begin{minipage}[t]{\linewidth}\raggedright

LGB Status

\end{minipage} & \begin{minipage}[t]{\linewidth}\raggedright

0.092***

\end{minipage} & \begin{minipage}[t]{\linewidth}\raggedright

0.277***

\end{minipage} & \begin{minipage}[t]{\linewidth}\raggedright

-0.379***

\end{minipage} & \begin{minipage}[t]{\linewidth}\raggedright

-0.335***

\end{minipage} & \begin{minipage}[t]{\linewidth}\raggedright

0.613***

\end{minipage} \\
& \begin{minipage}[t]{\linewidth}\raggedright

(0.017)

\end{minipage} & \begin{minipage}[t]{\linewidth}\raggedright

(0.005)

\end{minipage} & \begin{minipage}[t]{\linewidth}\raggedright

(0.026)

\end{minipage} & \begin{minipage}[t]{\linewidth}\raggedright

(0.025)

\end{minipage} & \begin{minipage}[t]{\linewidth}\raggedright

(0.039)

\end{minipage} \\
\begin{minipage}[t]{\linewidth}\raggedright

Political Affiliation

(Ref. = Conservative)

\end{minipage} & & & & & \\
\begin{minipage}[t]{\linewidth}\raggedright

Independent

\end{minipage} & \begin{minipage}[t]{\linewidth}\raggedright

-0.284***

(0.015)

\end{minipage} & \begin{minipage}[t]{\linewidth}\raggedright

-0.191***

(0.005)

\end{minipage} & \begin{minipage}[t]{\linewidth}\raggedright

0.011

(0.025)

\end{minipage} & \begin{minipage}[t]{\linewidth}\raggedright

-0.002

(0.025)

\end{minipage} & \begin{minipage}[t]{\linewidth}\raggedright

0.003

(0.025)

\end{minipage} \\
Liberal & \begin{minipage}[t]{\linewidth}\raggedright

-0.284***

(0.015)

\end{minipage} & \begin{minipage}[t]{\linewidth}\raggedright
-0.207***

(0.004)

\end{minipage} & \begin{minipage}[t]{\linewidth}\raggedright

-0.011

(0.022)

\end{minipage} & \begin{minipage}[t]{\linewidth}\raggedright

-0.035

(0.022)

\end{minipage} & \begin{minipage}[t]{\linewidth}\raggedright

-0.029

(0.022)

\end{minipage} \\
Other & \begin{minipage}[t]{\linewidth}\raggedright

-0.658***

(0.046)

\end{minipage} & \begin{minipage}[t]{\linewidth}\raggedright
-0.330***

(0.012)

\end{minipage} & \begin{minipage}[t]{\linewidth}\raggedright

-0.345***

(0.069)

\end{minipage} & \begin{minipage}[t]{\linewidth}\raggedright

-0.374***

(0.071)

\end{minipage} & \begin{minipage}[t]{\linewidth}\raggedright

-0.363***

(0.070)

\end{minipage} \\
Prefer not to say & \begin{minipage}[t]{\linewidth}\raggedright

0.083*

(0.034)

\end{minipage} & \begin{minipage}[t]{\linewidth}\raggedright
-0.143***

(0.010)

\end{minipage} & \begin{minipage}[t]{\linewidth}\raggedright

0.493***

(0.050)

\end{minipage} & \begin{minipage}[t]{\linewidth}\raggedright

0.483***

(0.050)

\end{minipage} & \begin{minipage}[t]{\linewidth}\raggedright

0.492***

(0.050)

\end{minipage} \\
\begin{minipage}[t]{\linewidth}\raggedright

Toxicity Rating

\end{minipage} & & & \begin{minipage}[t]{\linewidth}\raggedright

1.72***

\end{minipage} & \begin{minipage}[t]{\linewidth}\raggedright

1.90***

\end{minipage} & \begin{minipage}[t]{\linewidth}\raggedright

1.92***

\end{minipage} \\
& & & \begin{minipage}[t]{\linewidth}\raggedright

(0.013)

\end{minipage} & \begin{minipage}[t]{\linewidth}\raggedright

(0.012)

\end{minipage} & \begin{minipage}[t]{\linewidth}\raggedright

(0.012)

\end{minipage} \\
\begin{minipage}[t]{\linewidth}\raggedright

Female*Toxicity Rating

\end{minipage} & & & \begin{minipage}[t]{\linewidth}\raggedright

0.209***

\end{minipage} & & \\
& & & \begin{minipage}[t]{\linewidth}\raggedright

(0.014)

\end{minipage} & & \\
\begin{minipage}[t]{\linewidth}\raggedright

Asian*Toxicity Rating

\end{minipage} & & & & \begin{minipage}[t]{\linewidth}\raggedright

0.206***

\end{minipage} & \\
& & & & \begin{minipage}[t]{\linewidth}\raggedright

(0.034)

\end{minipage} & \\
\begin{minipage}[t]{\linewidth}\raggedright

Black*Toxicity Rating

\end{minipage} & & & & \begin{minipage}[t]{\linewidth}\raggedright

-0.499***

\end{minipage} & \\
& & & & \begin{minipage}[t]{\linewidth}\raggedright

(0.018)

\end{minipage} & \\
\begin{minipage}[t]{\linewidth}\raggedright

Hispanic*Toxicity Rating

\end{minipage} & & & & \begin{minipage}[t]{\linewidth}\raggedright

-0.051

\end{minipage} & \\
& & & & \begin{minipage}[t]{\linewidth}\raggedright

(0.045)

\end{minipage} & \\
\begin{minipage}[t]{\linewidth}\raggedright

Multiracial*Toxicity Rating

\end{minipage} & & & & \begin{minipage}[t]{\linewidth}\raggedright

0.217***

(0.041)

\end{minipage} & \\
\begin{minipage}[t]{\linewidth}\raggedright

Other Race*Toxicity Rating

\end{minipage} & & & & \begin{minipage}[t]{\linewidth}\raggedright

-0.261***

(0.051)

\end{minipage} & \\
\begin{minipage}[t]{\linewidth}\raggedright

LGB*Toxicity Rating

\end{minipage} & & & & & \begin{minipage}[t]{\linewidth}\raggedright

-0.474***

\end{minipage} \\
& & & & & \begin{minipage}[t]{\linewidth}\raggedright

(0.017)

\end{minipage} \\
\begin{minipage}[t]{\linewidth}\raggedright

Observations

\end{minipage} & \begin{minipage}[t]{\linewidth}\raggedright

236,645

\end{minipage} & \begin{minipage}[t]{\linewidth}\raggedright

501,540

\end{minipage} & \begin{minipage}[t]{\linewidth}\raggedright

236,645

\end{minipage} & \begin{minipage}[t]{\linewidth}\raggedright

236,645

\end{minipage} & \begin{minipage}[t]{\linewidth}\raggedright

236,645

\end{minipage} \\
\end{longtable}

\textbf{Notes:} (1) Robust standard errors are in parentheses (clustered at comment level). (2) Model 1 is used when plotting Figure 1 in main text. When estimating moderation demand, because conditional fixed-effects logistic regression is identified only by comments exhibiting variation in moderation judgments, comments for which all respondents made the same binary decision do not contribute to estimation. The moderation-demand models therefore use 236,645 ratings from 16,172 respondents across 47,553 comments. A total of 54,910 comments containing 264,895 ratings have unanimous outcomes and do not contribute to estimation. The excluded comments remain part of the descriptive analyses and simulations. Model 1 is used when plotting Figure 1. (3) Model 2 is used when plotting Figure S1 in SI. (4) Model 3-5 estimates moderation demand by including the interaction terms between respondent identity and comment toxicity level perceived by the respondent. They provide additional evidence that the stratified moderation demand across groups cannot be simply reduced to group differences in toxicity perception, by showing that different social groups react to what they consider as toxic content differently.

\textsuperscript{*} \emph{p}\textless{} 0.05, \textsuperscript{**} \emph{p} \textless{} 0.01, \textsuperscript{***} \emph{p}\textless{} 0.001

\clearpage
\noindent\textbf{Table S3. Decision Rules}\par\smallskip

\begin{longtable}[]{@{}
  >{\raggedright\arraybackslash}p{(\linewidth - 4\tabcolsep) * \real{0.3137}}
  >{\raggedright\arraybackslash}p{(\linewidth - 4\tabcolsep) * \real{0.2353}}
  >{\raggedright\arraybackslash}p{(\linewidth - 4\tabcolsep) * \real{0.4510}}@{}}
\toprule\noalign{}
\begin{minipage}[b]{\linewidth}\raggedright
\textbf{Decision rule}
\end{minipage} & \begin{minipage}[b]{\linewidth}\raggedright
\textbf{Panel size}
\end{minipage} & \begin{minipage}[b]{\linewidth}\raggedright
\textbf{Comment removed when...}
\end{minipage} \\
\midrule\noalign{}
\endhead
\bottomrule\noalign{}
\endlastfoot
One reviewer & 1 & assigned reviewer removes \\
Majority of 3 & 3 & at least 2 reviewers remove \\
Majority of 5 & 5 & at least 3 reviewers remove \\
Supermajority of 5 & 5 & at least 4 reviewers remove \\
Remove if any of 5 flags & 5 & at least 1 reviewer removes \\
Majority of 10 & 10 & at least 6 reviewers remove \\
Majority of 20 & 20 & at least 11 reviewers remove \\
\end{longtable}

\clearpage
\noindent\textbf{Table S4. Group-Specific Protection Gaps Under Alternative Decision Rules}\par\smallskip

\begin{longtable}[]{@{}
  >{\raggedright\arraybackslash}p{(\linewidth - 14\tabcolsep) * \real{0.1153}}
  >{\raggedright\arraybackslash}p{(\linewidth - 14\tabcolsep) * \real{0.1248}}
  >{\raggedright\arraybackslash}p{(\linewidth - 14\tabcolsep) * \real{0.1271}}
  >{\raggedright\arraybackslash}p{(\linewidth - 14\tabcolsep) * \real{0.1271}}
  >{\raggedright\arraybackslash}p{(\linewidth - 14\tabcolsep) * \real{0.1248}}
  >{\raggedright\arraybackslash}p{(\linewidth - 14\tabcolsep) * \real{0.1250}}
  >{\raggedright\arraybackslash}p{(\linewidth - 14\tabcolsep) * \real{0.1271}}
  >{\raggedright\arraybackslash}p{(\linewidth - 14\tabcolsep) * \real{0.1287}}@{}}
\toprule\noalign{}
\begin{minipage}[b]{\linewidth}\centering
\end{minipage} & \begin{minipage}[b]{\linewidth}\centering
\textbf{One reviewer}
\end{minipage} & \begin{minipage}[b]{\linewidth}\centering
\textbf{Majority of 3}
\end{minipage} & \begin{minipage}[b]{\linewidth}\centering
\textbf{Majority of 5}
\end{minipage} & \begin{minipage}[b]{\linewidth}\centering
\textbf{Super-majority of 5}
\end{minipage} & \begin{minipage}[b]{\linewidth}\centering
\textbf{Remove if any of 5 flags}
\end{minipage} & \begin{minipage}[b]{\linewidth}\centering
\textbf{Majority of 10}
\end{minipage} & \begin{minipage}[b]{\linewidth}\centering
\textbf{Majority of 20}
\end{minipage} \\
\midrule\noalign{}
\endhead
\bottomrule\noalign{}
\endlastfoot
Women $-$ Men & $-$3.7 $\rightarrow$ +8.3 (+12.0) & $-$3.3 $\rightarrow$ +8.3 (+11.5) & $-$3.0 $\rightarrow$ +8.1 (+11.1) & $-$1.2 $\rightarrow$ +3.9

(+5.1) & $-$11.9 $\rightarrow$ +21.1 (+33.0) & $-$2.1 $\rightarrow$ +6.6

(+8.7) & $-$2.2 $\rightarrow$ +7.1

(+9.3) \\
LGB $-$ Non-LGB & $-$5.9 $\rightarrow$ +4.8 (+10.7) & $-$6.0 $\rightarrow$ +3.6

(+9.6) & $-$5.9 $\rightarrow$ +2.9

(+8.9) & $-$2.7 $\rightarrow$ +0.8

(+3.6) & $-$13.8 $\rightarrow$ +18.6 (+32.4) & $-$4.7 $\rightarrow$ +1.5

(+6.2) & $-$5.1 $\rightarrow$ +1.4

(+6.5) \\
Black $-$ White & $-$5.4 $\rightarrow$ +6.0 (+11.4) & $-$5.7 $\rightarrow$ +5.3 (+11.0) & $-$5.7 $\rightarrow$ +4.8 (+10.5) & $-$2.8 $\rightarrow$ +1.7

(+4.5) & $-$11.7 $\rightarrow$ +20.5 (+32.1) & $-$4.7 $\rightarrow$ +3.2

(+7.9) & $-$5.1 $\rightarrow$ +3.3

(+8.4) \\
Asian $-$ White & +1.2 $\rightarrow$ +16.6 (+15.4) & +1.2 $\rightarrow$ +17.9 (+16.6) & +1.2 $\rightarrow$ +18.3 (+17.1) & +0.5 $\rightarrow$ +11.9 (+11.4) & +3.3 $\rightarrow$ +28.6 (+25.3) & +0.9 $\rightarrow$ +17.0 (+16.1) & +1.0 $\rightarrow$ +18.1 (+17.1) \\
Hispanic $-$ White & 0.0 $\rightarrow$ +13.9 (+13.9) & 0.0 $\rightarrow$ +14.7 (+14.7) & 0.0 $\rightarrow$ +15.0 (+15.0) & 0.0 $\rightarrow$

+8.8

(+8.8) & +0.3 $\rightarrow$ +26.0 (+25.7) & 0.0 $\rightarrow$ +13.6 (+13.6) & 0.0 $\rightarrow$ +14.7 (+14.7) \\
\end{longtable}

\textbf{Notes:} Each cell reports the toxicity-reduction gap when the focal group makes up 10\% of the moderator pool and when it makes up 90\% (change magnitude in parentheses). Values are percentage points. Gaps are defined as focal group minus comparison group. Positive changes indicate that increasing the focal group's representation shifts relative toxicity reduction toward that group.

\clearpage
\noindent\textbf{Table S5. Demographic Characteristics of Study Samples (\%)}\par\smallskip

\begin{longtable}[]{@{}
  >{\raggedright\arraybackslash}p{(\linewidth - 10\tabcolsep) * \real{0.2463}}
  >{\centering\arraybackslash}p{(\linewidth - 10\tabcolsep) * \real{0.1527}}
  >{\centering\arraybackslash}p{(\linewidth - 10\tabcolsep) * \real{0.1691}}
  >{\centering\arraybackslash}p{(\linewidth - 10\tabcolsep) * \real{0.1429}}
  >{\centering\arraybackslash}p{(\linewidth - 10\tabcolsep) * \real{0.1379}}
  >{\centering\arraybackslash}p{(\linewidth - 10\tabcolsep) * \real{0.1511}}@{}}
\toprule\noalign{}
\endhead
\bottomrule\noalign{}
\endlastfoot
& Toxicity Perspective & \multicolumn{2}{>{\centering\arraybackslash}p{(\linewidth - 10\tabcolsep) * \real{0.3120} + 2\tabcolsep}}{%
Moderator Survey} & GSS

2024 & Reddit Moderators \\
& & Moderation Experience & No Experience & & \\
\emph{Gender} & & & & & \\
Women & 53.59 & 47.6 & 51.14 & 55.41 & 24.37 \\
Men & 46.41 & 52.4 & 48.86 & 44.59 & 75.63 \\
\emph{Race} & & & & & \\
White & 72.17 & 78.17 & 78.92 & 61.71 & - \\
Black & 12.92 & 13.97 & 10.27 & 15.09 & - \\
Hispanic & 2.85 & 2.18 & 3.89 & 13.91 & - \\
Asian & 5.68 & 3.49 & 5.3 & 3.72 & - \\
Multiracial/Other & 6.37 & 2.18 & 1.62 & 5.57 & - \\
\emph{Education} & & & & & \\
Less than high school & 0.54 & - & 0.65 & 8.75 & 1.18 \\
High school & 9.09 & 13.1 & 14.16 & 24.41 & 8.40 \\
Some college or

associate's

degree & 31.35 & 33.19 & 35.89 & 27.91 & 39.50 \\
Bachelor's degree & 40.81 & 30.13 & 31.78 & 20.5 & 31.93 \\
Graduate or

professional

degree & 18.2 & 23.58 & 17.51 & 18.43 & 18.99 \\
\emph{Age} & & & & & \\
18 - 24 & 12.59 & 8.3 & 8.97 & 7.16 & 22.86 \\
25 - 34 & 40.83 & 25.76 & 17.19 & 15.88 & 45.71 \\
35 - 44 & 24.43 & 15.72 & 16.65 & 18.33 & 21.85 \\
45 - 54 & 12.53 & 20.09 & 15.78 & 14.86 & 6.72 \\
55 - 64 & 6.76 & 19.65 & 24.76 & 17.35 & 2.02 \\
65 or older & 2.85 & 10.48 & 16.65 & 26.42 & 0.84 \\
\emph{LGB Identity} & & & & & \\
LGB & 16.63 & 21.83 & 11.57 & 5.98 & - \\
Non-LGB & 83.37 & 78.17 & 88.43 & 94.01 & - \\
N of Participants

N of Comments

N of Individual Ratings & 16,221

102,463

501,540 & 230

-

- & 931

-

- & 3,142

-

- & 595

-

- \\
\end{longtable}

\textbf{Notes:} Race and LGB identity are not reported in published Reddit moderator data. The Reddit-matched simulation rakes to gender, age, and education margins only.

\clearpage
\noindent\textbf{Table S6. Changes in Relative Toxicity Reduction Under Empirical Benchmark Pools Compared with the Nationally Representative Pool, by Decision Rule.}\par\smallskip

\textbf{Panel A. Gender}

\begin{longtable}[]{@{}
  >{\raggedright\arraybackslash}p{(\linewidth - 8\tabcolsep) * \real{0.2115}}
  >{\raggedright\arraybackslash}p{(\linewidth - 8\tabcolsep) * \real{0.1058}}
  >{\raggedright\arraybackslash}p{(\linewidth - 8\tabcolsep) * \real{0.2115}}
  >{\raggedright\arraybackslash}p{(\linewidth - 8\tabcolsep) * \real{0.2376}}
  >{\raggedright\arraybackslash}p{(\linewidth - 8\tabcolsep) * \real{0.2335}}@{}}
\toprule\noalign{}
\begin{minipage}[b]{\linewidth}\centering
\textbf{Decision rule}
\end{minipage} & \begin{minipage}[b]{\linewidth}\centering
\textbf{Group}
\end{minipage} & \begin{minipage}[b]{\linewidth}\centering
\textbf{Moderation-Experience Minus General Population}
\end{minipage} & \begin{minipage}[b]{\linewidth}\centering
\textbf{No-Experience Minus General Population}
\end{minipage} & \begin{minipage}[b]{\linewidth}\centering
\textbf{Reddit-Matched Minus General Population}
\end{minipage} \\
\midrule\noalign{}
\endhead
\bottomrule\noalign{}
\endlastfoot
Single reviewer & Women & -1.8 & -1.1 & -4.5 \\
& Men & -0.3 & -0.3 & 0.3 \\
Majority of 3 & Women & -2.0 & -1.3 & -4.9 \\
& Men & -0.7 & -0.5 & -0.4 \\
Majority of 5 & Women & -2.2 & -1.4 & -5.0 \\
& Men & -0.9 & -0.6 & -0.8 \\
Supermajority of 5 & Women & -1.3 & -0.7 & -2.6 \\
& Men & -0.7 & -0.4 & -0.9 \\
Remove if any of 5 reviewers flags & Women & -3.6 & -2.5 & -10.7 \\
& Men & 1.1 & -0.3 & 3.7 \\
Majority of 10 & Women & -2.0 & -1.3 & -4.3 \\
& Men & -1.1 & -0.7 & -1.3 \\
Majority of 20 & Women & -2.3 & -1.4 & -4.7 \\
& Men & -1.2 & -0.8 & -1.6 \\
\end{longtable}

\textbf{Panel B. LGB Status}

\begin{longtable}[]{@{}
  >{\raggedright\arraybackslash}p{(\linewidth - 8\tabcolsep) * \real{0.2115}}
  >{\raggedright\arraybackslash}p{(\linewidth - 8\tabcolsep) * \real{0.1154}}
  >{\raggedright\arraybackslash}p{(\linewidth - 8\tabcolsep) * \real{0.2212}}
  >{\raggedright\arraybackslash}p{(\linewidth - 8\tabcolsep) * \real{0.2308}}
  >{\raggedright\arraybackslash}p{(\linewidth - 8\tabcolsep) * \real{0.2212}}@{}}
\toprule\noalign{}
\begin{minipage}[b]{\linewidth}\centering
\textbf{Decision rule}
\end{minipage} & \begin{minipage}[b]{\linewidth}\centering
\textbf{Group}
\end{minipage} & \begin{minipage}[b]{\linewidth}\centering
\textbf{Moderation-Experience Minus General Population}
\end{minipage} & \begin{minipage}[b]{\linewidth}\centering
\textbf{No-Experience Minus General Population}
\end{minipage} & \begin{minipage}[b]{\linewidth}\centering
\textbf{Reddit-Matched Minus General Population}
\end{minipage} \\
\midrule\noalign{}
\endhead
\bottomrule\noalign{}
\endlastfoot
Single reviewer & LGB & 0.9 & 0.0 & -0.3 \\
& Non-LGB & -1.6 & -0.9 & -2.6 \\
Majority of 3 & LGB & 0.4 & -0.2 & -0.9 \\
& Non-LGB & -1.8 & -1.1 & -3.1 \\
Majority of 5 & LGB & 0.1 & -0.3 & -1.2 \\
& Non-LGB & -2.0 & -1.2 & -3.4 \\
Supermajority of 5 & LGB & -0.2 & -0.2 & -0.9 \\
& Non-LGB & -1.2 & -0.7 & -1.9 \\
Remove if any of 5 reviewers flags & LGB & 4.9 & 0.5 & 1.7 \\
& Non-LGB & -3.0 & -2.0 & -5.3 \\
Majority of 10 & LGB & -0.2 & -0.4 & -1.4 \\
& Non-LGB & -1.9 & -1.1 & -3.2 \\
Majority of 20 & LGB & -0.3 & -0.5 & -1.6 \\
& Non-LGB & -2.1 & -1.2 & -3.5 \\
\end{longtable}

\textbf{Panel C. Race}

\begin{longtable}[]{@{}
  >{\raggedright\arraybackslash}p{(\linewidth - 8\tabcolsep) * \real{0.2115}}
  >{\raggedright\arraybackslash}p{(\linewidth - 8\tabcolsep) * \real{0.1096}}
  >{\raggedright\arraybackslash}p{(\linewidth - 8\tabcolsep) * \real{0.2077}}
  >{\raggedright\arraybackslash}p{(\linewidth - 8\tabcolsep) * \real{0.2490}}
  >{\raggedright\arraybackslash}p{(\linewidth - 8\tabcolsep) * \real{0.2221}}@{}}
\toprule\noalign{}
\begin{minipage}[b]{\linewidth}\centering
\textbf{Decision rule}
\end{minipage} & \begin{minipage}[b]{\linewidth}\centering
\textbf{Group}
\end{minipage} & \begin{minipage}[b]{\linewidth}\centering
\textbf{Moderator-Experience Minus General Population}
\end{minipage} & \begin{minipage}[b]{\linewidth}\centering
\textbf{No-Experience Pool Minus General Population}
\end{minipage} & \begin{minipage}[b]{\linewidth}\centering
\textbf{Reddit-Matched Minus General Population}
\end{minipage} \\
\midrule\noalign{}
\endhead
\bottomrule\noalign{}
\endlastfoot
Single reviewer & White & -1.0 & -0.4 & -2.2 \\
& Black & -0.8 & -1.2 & -1.4 \\
& Asian & -1.7 & -0.8 & -1.9 \\
& Hispanic & -3.3 & -2.5 & -4.0 \\
Majority of 3 & White & -1.3 & -0.7 & -2.8 \\
& Black & -1.0 & -1.3 & -1.8 \\
& Asian & -1.9 & -1.0 & -2.4 \\
& Hispanic & -3.4 & -2.6 & -4.4 \\
Majority of 5 & White & -1.4 & -0.8 & -3.0 \\
& Black & -1.2 & -1.3 & -2.0 \\
& Asian & -2.0 & -1.0 & -2.7 \\
& Hispanic & -3.4 & -2.6 & -4.5 \\
Supermajority of 5 & White & -1.0 & -0.5 & -1.8 \\
& Black & -0.7 & -0.6 & -1.2 \\
& Asian & -1.2 & -0.6 & -1.6 \\
& Hispanic & -1.8 & -1.3 & -2.4 \\
Remove if any of 5 reviewers flags & White & -0.9 & -0.5 & -4.0 \\
& Black & -0.5 & -3.4 & -1.9 \\
& Asian & -3.2 & -1.6 & -3.3 \\
& Hispanic & -7.9 & -6.7 & -9.3 \\
Majority of 10 & White & -1.5 & -0.8 & -2.9 \\
& Black & -1.2 & -1.1 & -2.0 \\
& Asian & -1.9 & -1.0 & -2.6 \\
& Hispanic & -3.0 & -2.2 & -4.1 \\
Majority of 20 & White & -1.7 & -0.9 & -3.3 \\
& Black & -1.3 & -1.2 & -2.2 \\
& Asian & -2.1 & -1.1 & -3.0 \\
& Hispanic & -3.3 & -2.4 & -4.5 \\
\end{longtable}

\textbf{Notes:} Cells are percentage-point changes in relative toxicity reduction, calculated as the empirical benchmark pool minus the nationally representative pool for the same group and decision rule. Negative values mean that the empirical benchmark pool produces less toxicity reduction for that group than the nationally representative pool; positive values mean more toxicity reduction.

\clearpage
\noindent\textbf{Table S7. Persistent LGB and Black Under-Protection Across Benchmark Pools and Decision Rules}\par\smallskip

\begin{longtable}[]{@{}
  >{\raggedright\arraybackslash}p{(\linewidth - 8\tabcolsep) * \real{0.2324}}
  >{\raggedright\arraybackslash}p{(\linewidth - 8\tabcolsep) * \real{0.2027}}
  >{\raggedright\arraybackslash}p{(\linewidth - 8\tabcolsep) * \real{0.1791}}
  >{\raggedright\arraybackslash}p{(\linewidth - 8\tabcolsep) * \real{0.1761}}
  >{\raggedright\arraybackslash}p{(\linewidth - 8\tabcolsep) * \real{0.2098}}@{}}
\toprule\noalign{}
\begin{minipage}[b]{\linewidth}\centering
\textbf{Decision rule}
\end{minipage} & \begin{minipage}[b]{\linewidth}\centering
\textbf{LGB Minus Non-LGB}
\end{minipage} & \begin{minipage}[b]{\linewidth}\centering
\textbf{Black Minus White}
\end{minipage} & \begin{minipage}[b]{\linewidth}\centering
\textbf{Black Minus Asian}
\end{minipage} & \begin{minipage}[b]{\linewidth}\centering
\textbf{Black Minus Hispanic}
\end{minipage} \\
\midrule\noalign{}
\endhead
\bottomrule\noalign{}
\endlastfoot
Single reviewer & $-$6.4 to $-$3.9 & $-$5.4 to $-$3.8 & $-$5.1 to $-$3.8 & $-$5.4 to $-$2.7 \\
Majority of 3 & $-$6.5 to $-$4.2 & $-$5.7 to $-$4.1 & $-$5.4 to $-$4.2 & $-$5.7 to $-$3.1 \\
Majority of 5 & $-$6.4 to $-$4.2 & $-$5.6 to $-$4.1 & $-$5.4 to $-$4.3 & $-$5.8 to $-$3.3 \\
Supermajority of 5 & $-$3.0 to $-$1.9 & $-$2.7 to $-$1.9 & $-$2.5 to $-$2.0 & $-$2.8 to $-$1.6 \\
Remove if any of 5 flags & $-$15.6 to $-$7.6 & $-$12.0 to $-$6.9 & $-$11.0 to $-$6.4 & $-$11.3 to $-$3.9 \\
Majority of 10 & $-$5.1 to $-$3.3 & $-$4.6 to $-$3.4 & $-$4.3 to $-$3.5 & $-$4.8 to $-$2.7 \\
Majority of 20 & $-$5.4 to $-$3.5 & $-$4.9 to $-$3.6 & $-$4.6 to $-$3.8 & $-$5.2 to $-$2.9 \\
\end{longtable}

\textbf{Notes:} Values are percentage-point gaps in relative toxicity reduction. For each decision rule, ranges summarize the minimum and maximum gaps across the four benchmark pools. Negative values represent that the focal group receives less toxicity reduction than the reference group.

\clearpage

\noindent\textbf{Table S8. Full Survey Instrument for the Moderation-Experience Survey}\par\smallskip

\begin{longtable}[]{@{}
  >{\raggedright\arraybackslash}p{(\linewidth - 0\tabcolsep) * \real{1.0000}}@{}}
\toprule\noalign{}
\begin{minipage}[b]{\linewidth}
\raggedright

\textbf{Moderation Experience}

Content moderation involves reviewing or managing user-generated content on online platforms

(e.g., social media, forums, websites) to ensure it follows certain guidelines (e.g., removing harmful or inappropriate posts). Have you ever performed any content moderation tasks on any online platform (e.g., as a volunteer, community moderator, employee, etc.)?

- Yes

- No

- Not sure

\textbf{Demographic Questions}

In what year were you born?

- Year of birth

Which of the following best describes your race or ethnicity? Check all that apply.

- White

- Black

- Hispanic

- Asian

- Other

- Prefer not to answer

Which of the following best describes your gender?

- Woman

- Man

- Nonbinary

- Prefer not to answer

Which of the following best describes your sexual orientation?

- Heterosexual / straight

- Bisexual

- Gay or lesbian

- Other

- Prefer not to answer

What is the highest level of school you have completed or the highest degree you have received?

- Less than high school

- High school graduate

- Some college

- 2 year degree (e.g., Associate\textquotesingle s degree or vocational training)

- 4 year degree (e.g., Bachelor\textquotesingle s degree)

- Professional / master\textquotesingle s degree

- Doctorate

- Prefer not to answer
\end{minipage} \\
\midrule\noalign{}
\endhead

\endlastfoot
\end{longtable}

\clearpage
\noindent\textbf{Table S9. Observed Rating Coverage by Group}\par\smallskip

\begin{longtable}[]{@{}
  >{\raggedright\arraybackslash}p{(\linewidth - 10\tabcolsep) * \real{0.1029}}
  >{\raggedright\arraybackslash}p{(\linewidth - 10\tabcolsep) * \real{0.1460}}
  >{\raggedright\arraybackslash}p{(\linewidth - 10\tabcolsep) * \real{0.1674}}
  >{\raggedright\arraybackslash}p{(\linewidth - 10\tabcolsep) * \real{0.1921}}
  >{\raggedright\arraybackslash}p{(\linewidth - 10\tabcolsep) * \real{0.1950}}
  >{\raggedright\arraybackslash}p{(\linewidth - 10\tabcolsep) * \real{0.1966}}@{}}
\toprule\noalign{}
\begin{minipage}[b]{\linewidth}\centering
\textbf{Group}
\end{minipage} & \begin{minipage}[b]{\linewidth}\centering
\textbf{Respondents}
\end{minipage} & \begin{minipage}[b]{\linewidth}\centering
\textbf{Individual rating records}
\end{minipage} & \begin{minipage}[b]{\linewidth}\centering
\textbf{Comments with $\geq$1 group rating}
\end{minipage} & \begin{minipage}[b]{\linewidth}\centering
\textbf{Comments with $\geq$2 group ratings}
\end{minipage} & \begin{minipage}[b]{\linewidth}\centering
\textbf{Comments with $\geq$3 group ratings}
\end{minipage} \\
\midrule\noalign{}
\endhead
\bottomrule\noalign{}
\endlastfoot
Women & 8,693 & 263,540 & 97.0\% & 82.4\% & 51.6\% \\
Men & 7,528 & 238,000 & 95.2\% & 76.0\% & 42.3\% \\
LGB & 2,698 & 78,800 & 55.6\% & 17.4\% & 3.0\% \\
Non-LGB & 13,523 & 422,740 & 99.9\% & 99.5\% & 95.0\% \\
White & 11,707 & 364,000 & 99.8\% & 97.0\% & 83.9\% \\
Black & 2,096 & 63,120 & 47.9\% & 11.7\% & 1.5\% \\
Asian & 922 & 30,760 & 26.5\% & 3.1\% & 0.3\% \\
Hispanic & 462 & 13,220 & 12.0\% & 0.8\% & 0.0\% \\
\end{longtable}

\textbf{Notes:} Each individual rating record contains both a toxicity rating and a moderation decision for the same respondent-comment pair, so observed coverage is identical for group-specific toxicity ratings and moderation decisions. The table reports the share of all 102,463 comments with at least one, two, or three observed ratings from each group. Coverage is uneven across groups. Women, men, White respondents, and non-LGB respondents rated nearly all comments, whereas coverage is substantially thinner for LGB respondents and especially for Black, Asian, and Hispanic respondents. The alternative decision-rule simulation imputes missing moderation decisions but does not impute missing toxicity ratings; toxicity-reduction outcomes therefore use observed group-specific toxicity ratings only.

\clearpage
\noindent\textbf{Table S10. Raking Diagnostics for Empirically Grounded Simulation Scenarios Based on Observed Removal Decisions.}\par\smallskip

\begin{longtable}[]{@{}
  >{\raggedright\arraybackslash}p{(\linewidth - 4\tabcolsep) * \real{0.2748}}
  >{\raggedleft\arraybackslash}p{(\linewidth - 4\tabcolsep) * \real{0.3533}}
  >{\raggedleft\arraybackslash}p{(\linewidth - 4\tabcolsep) * \real{0.3719}}@{}}
\toprule\noalign{}
\endhead
\bottomrule\noalign{}
\endlastfoot
\begin{minipage}[t]{\linewidth}\raggedright

\textbf{Scenario}

\end{minipage} & \begin{minipage}[t]{\linewidth}\raggedleft

\textbf{Max Absolute Error}

\end{minipage} & \begin{minipage}[t]{\linewidth}\raggedleft

\textbf{Mean Absolute Error}

\end{minipage} \\
\begin{minipage}[t]{\linewidth}\raggedright

General-Population

\end{minipage} & \begin{minipage}[t]{\linewidth}\raggedleft

0.0272

\end{minipage} & \begin{minipage}[t]{\linewidth}\raggedleft

0.0032

\end{minipage} \\
\begin{minipage}[t]{\linewidth}\raggedright

Moderation-Experience

\end{minipage} & \begin{minipage}[t]{\linewidth}\raggedleft

0.0254

\end{minipage} & \begin{minipage}[t]{\linewidth}\raggedleft

0.0017

\end{minipage} \\
\begin{minipage}[t]{\linewidth}\raggedright

No-Experience

\end{minipage} & \begin{minipage}[t]{\linewidth}\raggedleft

0.0236

\end{minipage} & \begin{minipage}[t]{\linewidth}\raggedleft

0.0015

\end{minipage} \\
\begin{minipage}[t]{\linewidth}\raggedright

Reddit-Matched

\end{minipage} & \begin{minipage}[t]{\linewidth}\raggedleft

0.0000

\end{minipage} & \begin{minipage}[t]{\linewidth}\raggedleft

0.0000

\end{minipage} \\
\end{longtable}

\textbf{Notes:} This table presents the maximum and mean absolute errors between the empirical target margins of our four reference populations and the achieved margins in our weighted dataset following the raking procedure. Errors are expressed as decimal proportions. The Reddit-matched scenario rakes only to the available gender, age, and education margins, which the procedure matches exactly.

\clearpage
\noindent\textbf{Table S11. Raking and Sampling Accuracy for Alternative Decision-Rule Simulations Based on Imputed Decisions.}\par\smallskip

\begin{longtable}[]{@{}
  >{\raggedright\arraybackslash}p{(\linewidth - 10\tabcolsep) * \real{0.1576}}
  >{\raggedright\arraybackslash}p{(\linewidth - 10\tabcolsep) * \real{0.1178}}
  >{\raggedright\arraybackslash}p{(\linewidth - 10\tabcolsep) * \real{0.1788}}
  >{\raggedright\arraybackslash}p{(\linewidth - 10\tabcolsep) * \real{0.1709}}
  >{\raggedright\arraybackslash}p{(\linewidth - 10\tabcolsep) * \real{0.1906}}
  >{\raggedright\arraybackslash}p{(\linewidth - 10\tabcolsep) * \real{0.1842}}@{}}
\toprule\noalign{}
\begin{minipage}[b]{\linewidth}\centering
\textbf{Scenario set}
\end{minipage} & \begin{minipage}[b]{\linewidth}\centering
\textbf{No. scenarios}
\end{minipage} & \begin{minipage}[b]{\linewidth}\centering
\textbf{Median absolute error in raked margins}
\end{minipage} & \begin{minipage}[b]{\linewidth}\centering
\textbf{Max absolute error in raked margins}
\end{minipage} & \begin{minipage}[b]{\linewidth}\centering
\textbf{Median absolute error in sampled pool share}
\end{minipage} & \begin{minipage}[b]{\linewidth}\centering
\textbf{Max absolute error in sampled pool share}
\end{minipage} \\
\midrule\noalign{}
\endhead
\bottomrule\noalign{}
\endlastfoot
Benchmark pools & 4 & 0.000 & 0.087 & 0.08 & 0.49 \\
Female share scenarios & 9 & 0.000 & 0.120 & 0.11 & 0.32 \\
LGB share scenarios & 9 & 0.000 & 0.514 & 0.13 & 0.46 \\
White share scenarios & 9 & 0.000 & 0.227 & 0.17 & 0.38 \\
Black share scenarios & 9 & 0.000 & 0.563 & 0.18 & 0.57 \\
Asian share scenarios & 9 & 0.000 & 3.686 & 0.26 & 2.93 \\
Hispanic share scenarios & 9 & 0.000 & 0.927 & 0.24 & 1.20 \\
\end{longtable}

\textbf{Notes:} Values are percentage-point absolute errors. Raked-margin errors compare the target demographic margins with the achieve weighted margins after raking. Sampled-pool errors compare the intended focal-group share with the mean achieved share across 2000 sampled 20-person moderator pools. Benchmark pools refer to the empirical grounded moderator-pool scenarios; group-share scenarios refer to counterfactual scenarios varying the focal group's share from 10\% to 90\%. As in the observed-data simulation, raking weights were capped at 10 to avoid assign extreme influence to sparse demographic cells. Overall, the diagnostics indicate that the raking and sampling procedures closely reproduce the intended moderator-pool compositions; the largest deviations occur in the Asian-share scenarios, where sparse intersecting demographic cells make exact multidimensional balancing more difficult.

\end{document}